\documentclass[lettersize,journal]{IEEEtran}  
\usepackage{amsmath,amsfonts}
\usepackage{algorithmic}
\usepackage{algorithm}
\usepackage{array}
\usepackage[caption=false,font=normalsize,labelfont=sf,textfont=sf]{subfig}
\usepackage{textcomp}
\usepackage{stfloats}
\usepackage{url}
\usepackage{verbatim}
\usepackage{graphicx}
\usepackage{multirow}
\usepackage{tabularx}
\usepackage{cite}

\begin{document}

\title{DeFTAN-II: Efficient Multichannel Speech Enhancement with Subgroup Processing}

\author{Dongheon Lee, and Jung-Woo Choi, \textit{Member, IEEE}}
        % <-this % stops a space
% \thanks{Manuscript received June 19, 2021; revised August 16, 2021.}}

\markboth{Journal of \LaTeX\ Class Files,~Vol.~14, No.~8, June~2023}%
{Shell \MakeLowercase{\textit{et al.}}: A Sample Article Using IEEEtran.cls for IEEE Journals}

% \IEEEpubid{0000--0000/00\$00.00~\copyright~2023 IEEE}

\maketitle

\begin{abstract}
In this work, we present DeFTAN-II, an efficient multichannel speech enhancement model based on transformer architecture and subgroup processing. 
Despite the success of transformers in speech enhancement, they face challenges in capturing local relations, reducing the high computational complexity, and lowering memory usage. To address these limitations, we introduce subgroup processing in our model, combining subgroups of locally emphasized features with other subgroups containing original features. The subgroup processing is implemented in several blocks of the proposed network. In the proposed split dense blocks extracting spatial features, a pair of subgroups is sequentially concatenated and processed by convolution layers to effectively reduce the computational complexity and memory usage. For the F- and T-transformers extracting temporal and spectral relations, we introduce cross-attention between subgroups to identify relationships between locally emphasized and non-emphasized features. The dual-path feedforward network then aggregates attended features in terms of the gating of local features processed by dilated convolutions. Through extensive comparisons with state-of-the-art multichannel speech enhancement models, we demonstrate that DeFTAN-II with subgroup processing outperforms existing methods at significantly lower computational complexity. Moreover, we evaluate the model's generalization capability on real-world data without fine-tuning, which further demonstrates its effectiveness in practical scenarios.
\end{abstract}


\section{Introduction}
\IEEEPARstart{I}{n} recent years, speech enhancement has gained substantial interest in several real-world applications, including augmented reality/virtual reality \cite{donley2021easycom, guiraud2022introduction}, teleconferencing, and noise-canceling headphones.
Among the different research fields involved with speech enhancement, multichannel speech enhancement has attracted particular attention due to its ability to leverage spatial information and superior performance compared to single-channel methods \cite{yoshioka2015ntt}. The primary objective of multichannel speech enhancement is to restore clean speech by reducing noise and reverberation from measured multichannel speech. While it is possible to design multi-input/multi-output models \cite{ren2021causal, pandey2022tparn} for restoring multichannel clean speech, the majority of approaches focus on multi-input/single-output models \cite{tolooshams2020channel, wang2020complex, wang2020multi, tesch2022insights, li2022embedding, pandey2022multichannel, liu2022drc, shubo2023spatial, yang2023mcnet, lee2023deft} that aim to restore the clean speech of a reference channel.

Speech enhancement has been developed alongside speech separation. Early research focused on estimating only the magnitude of the short-time Fourier transform (STFT) of clean speech \cite{xu2013experimental, lu2013speech, xu2014regression, isik2016single}. However, since estimating phase information is also crucial, studies \cite{williamson2015complex, park2017fully, wang2018supervised} have been conducted to address phase estimation. However, the limitation of phase reconstruction remained, leading to the development of time-domain end-to-end techniques that employ the waveform as input and output the estimated waveform \cite{fu2017raw, luo2018tasnet, venkataramani2018end, luo2019conv}. In particular, the convolutional time-domain audio separation network (Conv-TasNet) \cite{luo2019conv} introduced a learnable short encoder-decoder kernel for mask estimation, which has been widely adopted in subsequent studies \cite{rixen2022sfsrnet, tzinis2020sudo, rixen2022qdpn} for both speech enhancement and separation. Additionally, the dual-path recurrent neural network \cite{luo2020dual} introduced time-domain chunking, where features are split into intra-/inter-chunks and dual-path processing that iteratively processes each chunk during training.

In addition to the time-domain end-to-end models, alternative approaches have been proposed for estimating the complex spectrogram of clean speech using complex spectral mapping and masking techniques \cite{fu2017complex, tan2019complex, wang2020complex, wang2020multi}. Complex spectral mapping aims to reconstruct the input spectrogram into the spectrogram of clean speech using deep neural networks (DNNs). Moreover, complex spectral masking involves generating complex masks through element-wise multiplication between the input spectrogram and a mask, enabling the estimation of the spectrogram of clean speech. Estimating the complex spectrogram resolves the long-standing phase reconstruction issue in STFT-based approaches \cite{wang2020complex}. Furthermore, the importance of frequency domain information in speech enhancement has been rediscovered \cite{tesch2022insights}, leading to the development of loss functions that directly incorporate frequency information \cite{pandey2020densely, wang2020complex, yu2022dual, pandey2021dense} or directly handle frequency information using complex spectrogram as input of the DNNs \cite{wang2020complex, wang2020multi, yang2022tfpsnet, wang2023tf}.
% Currently, there is significant research focusing on applying the dual-path processing, initially proposed in the time domain, to the STFT domain, allowing separate processing in both the time and frequency domains [**].

Three main network architectures can be utilized for speech enhancement: recurrent neural network (RNN) \cite{elman1990finding} convolutional neural network (CNN) \cite{lecun1998gradient}, and transformer \cite{vaswani2017attention}. Early studies \cite{chen2015speech, weninger2015speech} focused on processing sequential information embedded in time-domain signals using RNNs, such as bidirectional long short-term memory (BLSTM) \cite{hochreiter1997long}. However, after the introduction of Conv-TasNet and the temporal convolutional network, dilated convolutions also gained significant popularity to achieve a wide receptive field. The separation transformer (SepFormer) \cite{subakan2021attention} utilized transformers, a widely used architecture in natural language processing, for constructing the entire mask estimation network. 
Since SepFormer, several state-of-the-art speech enhancement and separation models have utilized transformers in network design \cite{chen2020dual, yang2022tfpsnet, yu2022dual, yang2022tfpsnet, lee2023deft}. For multichannel speech enhancement, a triple-path attentive recurrent network (TPARN) \cite{pandey2022tparn} utilizes spatial, long-term, and short-term temporal relations using an attentive recurrent network (ARN). Pandey et al. \cite{pandey2022multichannel} introduced an attentive dense convolutional network (ADCN) that utilizes attention for capturing time-domain relationships. However, the above models only apply attention to the time domain, and it is difficult to aggregate the global relations in the frequency domain. 
Recently, dense frequency-time attentive network (DeFTAN) \cite{lee2023deft} has exhibited high performance by incorporating channel relationships with time and frequency relationships. The model utilizes DenseNet \cite{huang2017densely} to account for the spatial relation between channels, as well as transformer blocks extracting features in time and frequency dimensions. However, DenseNet sequentially stacks multichannel data, which increases computational complexity and memory usage. Another recently developed model, TF-GridNet \cite{wang2023tf}, employs BLSTMs in intra-frame full-band modules and sub-band temporal modules for capturing frequency and time information, respectively. TF-GridNet also employs cross-frame self-attention modules for aggregating time-domain relationships. This model has been extended to the multichannel speech enhancement model \cite{wang2022tf} but involves high computational complexity.

The transformer architecture excels in capturing global relations, but struggles to capture local relations, as it focuses on understanding the relationships across the entire sequence \cite{gulati2020conformer, guo2022cmt}. To address this, efforts have been made to capture local relations, e.g., using window attention such as in the swin (shifted windows)-transformer \cite{liu2021swin}, or using convolution on attention such as in convolutional self-attention (CSA) \cite{li2019enhancing}, conformer \cite{gulati2020conformer}, and CNNs meet transformers (CMT) \cite{guo2022cmt}. However, the swin-transformer captures local-global relations sequentially, because the window length sequentially decreases as the transformer block index increases. CSA emphasizes local relations of features using convolutions and finds their global relation, and thus, it can also be considered as sequential local-global processing. 

Another downside of the transformer architectures is their high computational complexity compared to CNNs \cite{vaswani2017attention}, especially due to the self-attention, where the complexity is proportional to the square of the sequence length \cite{vaswani2017attention}. Furthermore, while increasing the channel dimension improves the model's performance, it also leads to larger feature map sizes, increased computational complexity, and memory usage.
Therefore, it is necessary to efficiently extract features within a limited channel dimension while maintaining the sequence information.

In this study, we propose an efficient multichannel speech enhancement network, dense frequency-time attentive network II (DeFTAN-II), that can resolve the above issues of the transformer and DenseNet architectures \cite{huang2017densely} using subgroup processing. Subgroup processing is a method that utilizes both locally emphasized features and original features. Unlike the former model (DeFTAN) based on complex spectral masking, the proposed model adopts the complex spectral mapping approach \cite{wang2020complex, wang2020multi} that directly synthesizes STFT of clean speech. The contributions of this work are as follows:

First, we leverage the strengths of both the CNN and transformer architectures by employing subgroup processing at various stages, including attention layers, feedforward networks (FFWs), and dense blocks. Unlike CSA, which uses self-attention between locally emphasized features \cite{li2019enhancing}, we introduce cross-attention between asymmetric features obtained through subgroup processing, called convolutional efficient attention (CEA). By constructing two subgroups containing original features and locally emphasized features using CNNs, we demonstrate that cross-attention between these two subgroups leads to significant improvement in speech enhancement performance. Moreover, we adopt subgroup processing in the FFW to aggregate features, wherein dilated convolution is applied only to half of the features and then fused with the other features. This approach, referred to as the dual-path feedforward network (DPFN), balances computational complexity and performance, providing an effective and efficient solution for feature aggregation.

To enhance the computational efficiency, we introduce a novel dense block called a split dense block (SDB) to efficiently extract features within a limited channel dimension. The proposed SDB compresses features through channel splitting into subgroups, followed by sequential concatenation and convolution between subgroup features. This compression process yields improved performance without a subsequent increase of feature maps, which differentiates it from DenseNet \cite{huang2017densely}. Overall, our proposed model effectively combines the advantages of CNNs and transformers, offering enhanced performance while managing computational efficiency.

We demonstrate the improved performance, low data bias, and environment bias of the proposed model through various simulated datasets. Performance comparison with state-of-the-art models on three datasets (spatialized WSJCAM0 \cite{wang2020multi}, spatialized DNS challenge \cite{pandey2022tparn}, and L3DAS22 \cite{guizzo2022l3das22}) confirms that our proposed model has lower computational complexity and higher performance enhancement. To further investigate the real-world applicability and scalability of our model, we conduct experiments on real noisy and reverberant speech recorded in an office environment. Remarkably, we utilize a model pre-trained solely on simulated datasets, without any fine-tuning on real data. The results highlight the real-world feasibility of our model, suggesting its potential for practical deployment in challenging environments.

\section{Model Architecture}
The goal of multichannel speech enhancement can be expressed as follows. For an $M$ channel microphone array measuring $N$ samples, the relationship between the multichannel noisy reverberant speech, $\mathbf{y}\in\mathbb{R}^{M\times N}$, and clean speech, $\mathbf{s}\in\mathbb{R}^{M\times N}$, is given by

\begin{align}
\begin{aligned}
\mathbf{y}[n] &= \mathbf{x}[n] + \mathbf{v}[n]\\
     &= \mathbf{s}[n] * (1 + \mathbf{r}[n]) + \mathbf{v}[n],
\end{aligned}
\end{align}
where $\mathbf{x}[n]$, $\mathbf{v}[n]$, and $\mathbf{r}[n]$ %and $\mathbf{s}[n]$ 
represent the $n$ th sample of the reverberant speech, reverberant noise, and room impulse response (RIR) excluding the direct path, respectively. We aim to restore the reference channel signal $\mathbf{s}_{\mathrm{ref}}\in\mathbb{R}^{1\times N}$, where $\mathbf{s}=[\mathbf{s}_1, \mathbf{s}_2, \cdots, \mathbf{s}_{\mathrm{ref}}, \cdots, \mathbf{s}_M]$, from the measured noisy reverberant speech $\mathbf{y}$.

\begin{figure*}[!t]
\centering
\includegraphics[width=7in]{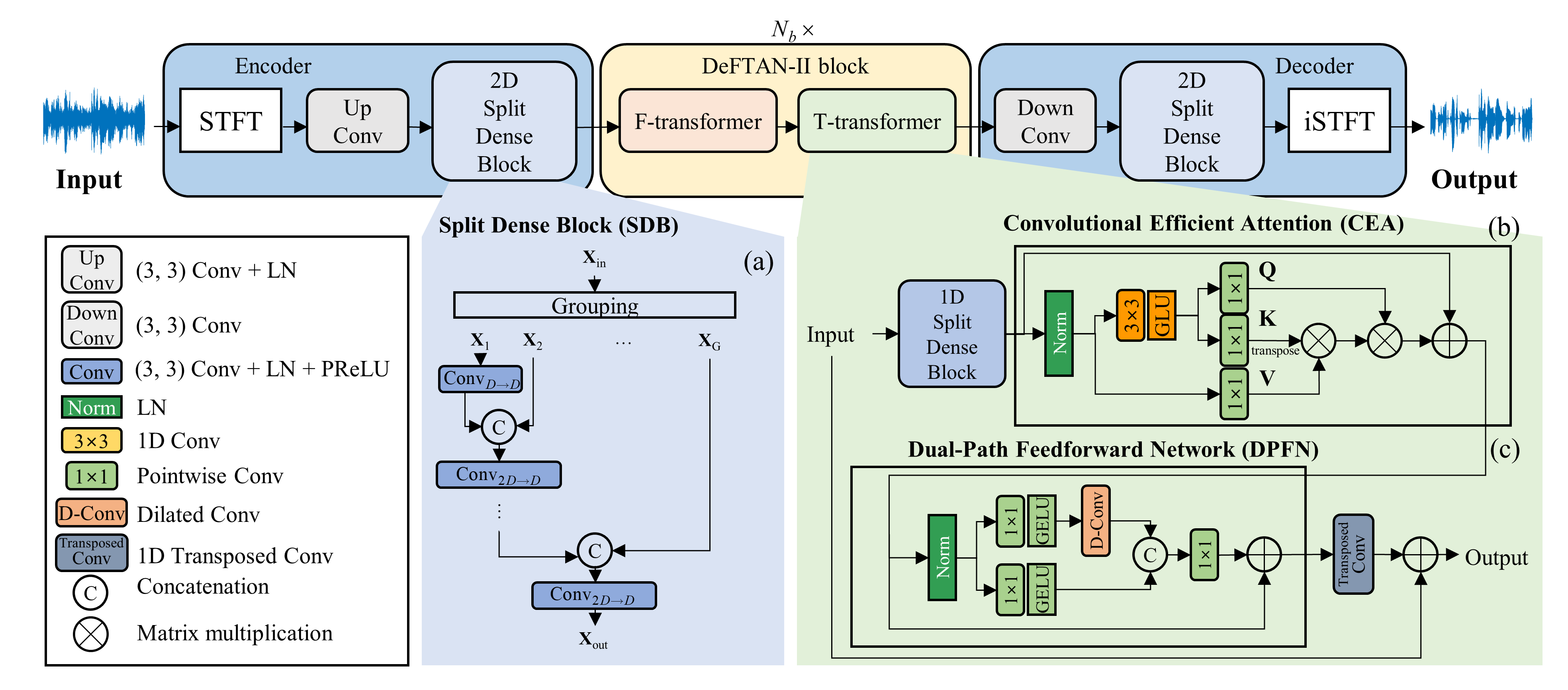}
\caption{Overall architecture of DeFTAN-II (a) split dense block (SDB) (b) convolutional efficient attention (CEA) (c) dual-path feedforward network (DPFN)}
\label{fig:overall_arch}
\end{figure*}

The overall architecture of DeFTAN-II is illustrated in Fig.\,\ref{fig:overall_arch}. DeFTAN-II is a non-causal complex spectral mapping model comprising three main stages: encoding, main processing, and decoding. 

%Before encoding, we normalize the sample variance of each input mixture, which has been known to improve the enhancement performance [**]. 
The encoding stage transforms the multichannel waveform into a multichannel complex spectrogram and extracts features before the main processing. 
The main-processing stage employs repeated DeFTAN-II blocks comprising an F-transformer and T-transformer that identify the local-global frequency and time relationships, respectively. In the decoding stage, the extracted features are combined to estimate the complex spectrogram of clean speech. The estimated spectrogram is then transformed into the time-domain waveform by applying the inverse STFT (iSTFT). The details of each stage are provided below.
%real and imaginary (RI) components of the estimated STFT. 

\subsection{Encoding}
Encoding is the initial stage of complex spectral mapping, where the time-domain waveform is transformed into the STFT domain. This step also involves increasing the channel dimension to extract a larger number of features. Before encoding, we normalize the sample variance of each input mixture, which has been known to improve the performance \cite{wang2022tf}. Then, performing a complex spectrogram on the $M$-channel time-domain waveform yields the real and imaginary (RI) components of size $T\times F$ for each channel. A Hamming window is applied as the window function for STFT, and only the single-sided spectra are taken as inputs. By stacking the RI components along the channel dimension, an input tensor of size $2M \times T\times F$ is obtained. To capture and enrich the local time-frequency (TF) features, the channel dimension is increased from $2M$ to $C$ using an up-convolution (Conv) layer including the 2D Conv of a $(3\times 3)$ kernel size and layer normalization (LN). The resulting tensor of size $C \times T\times F$ passes through the split dense block (SDB).

\subsubsection{2D Split Dense Block (2D SDB)}
The SDB is a dense block modified to efficiently extract inter-channel and local TF relationships. It is designed to address the computational complexity and memory usage issue of conventional dense blocks that exhibit a linear increase in the feature map size with respect to the number of dense layers. The SDB alleviates the complexity issue by splitting features and sequentially concatenating and shrinking them in channel dimension, as illustrated in Fig.\,\ref{fig:overall_arch}(a). First, the features are split into $G$ subgroups of features along the channel dimension, denoted as $\mathbf{X}=[\mathbf{X}_1, \mathbf{X}_2, \cdots, \mathbf{X}_g, \cdots, \mathbf{X}_G]\in\mathbb{R}^{C\times T\times F}$. 
%compresses the channel dimension through  $C$ into $D=\frac{C}{G}$, where $G$ is the number of feature groups.
The first subgroup $\mathbf{X}_1 \in \mathbb{R}^{D \times T\times F}$ is processed by %a dense layer
$(3\times 3)$ 2D Conv, followed by LN and the parametric rectified linear unit (PReLU) activation function \cite{he2015delving}. Here, $D=C/G$ is the size of the split channel dimension. The output of the same size is concatenated with the next subgroup $\mathbf{X}_{2}$ to serve as the input for the subsequent convolution layer. Except for the first layer, the subsequent layers of SDB take inputs of size $2D \times T\times F$ and output tensors of size $D \times T\times F$. This process is repeated to sequentially develop features from different subgroups. The sequential concatenation allows for the reutilization of previous features without a linear increase of channel dimension. Accordingly, the computational complexity and memory usage can be saved as compared to the conventional dense block \cite{pandey2022multichannel, lee2023deft} or grouped dense block. A group dense block is a dense block comprising group convolutions.% \cite{lee2023rt}.

To demonstrate the benefit of the SDB in computational complexity and memory usage, both two metrics of the dense block \cite{lee2023deft}, grouped dense block, and 2D SDB are compared in Table\,\ref{tab:DenseCost} for the convolution kernel size $k$. The computational complexity and memory usage are compared to design an efficient neural network \cite{hassani2023neighborhood}. The SDBs have the lowest values for both metrics, indicating that they have an efficient structure in terms of computational complexity and memory usage. In this work, $G$ is set to 4, and as a result, the memory usage of the 2D SDB is approximately four times lower than that of the dense and grouped dense blocks. The complexity analysis of subgroup processing is explained in the Appendix. The overall process of the SDB is presented in the algorithm\,\ref{alg:algSDB}.

The SDB is utilized in the encoder and decoder, as well as in the main processing. However, for the main processing, 1D convolutions along the frequency or time dimension are used, which are denoted as 1D SDBs to differentiate from 2D SDBs used in the encoder and decoder. The complexity of the 1D SDB is also presented in Table\,\ref{tab:DenseCost}, where $L$ denotes the length of the sequence dimension ($T$ or $F$), which has lower computational complexity compared to the 2D SDB. 

\begin{algorithm}[H]
\caption{2D Split Dense Block}\label{alg:algSDB}
\begin{algorithmic}
\STATE 
\STATE $[\mathbf{X}_1, \mathbf{X}_2, \cdots, \mathbf{X}_g, \cdots, \mathbf{X}_G]=\mathbf{X}$
\STATE $\mathbf{X}_{in}=\mathrm{Conv}_{D\rightarrow D}(\mathbf{X}_1)$
\STATE $g=1$
\STATE $\textbf{while }g<G$
\STATE \hspace{0.5cm}$\mathbf{X}_{out}=\mathrm{Conv}_{2D\rightarrow D}(\mathrm{Concat}([\mathbf{X}_{in}, \mathbf{X}_{g+1}]))$
\STATE \hspace{0.5cm}$\mathbf{X}_{in}=\mathbf{X}_{out}$
\STATE \hspace{0.5cm}$g=g+1$
\STATE \textbf{return} $\mathbf{X}_{out}$
\end{algorithmic}
\label{alg1}
\end{algorithm}

\begin{table}[!t]
\caption{Comparison of Complexity with Various Dense Blocks \label{tab:DenseCost}}
\renewcommand{\arraystretch}{1.5}
\centering
\begin{tabular}{ccc}
\hline
& Computational complexity & Memory usage \\
\hline\hline
Dense block & $\frac{G(G+1)}{2}C^2k^2TF$ & $\frac{G(G+1)}{2}C^2k^2$\\
\hline
Grouped dense block & $\frac{G+1}{2}C^2k^2TF$ & $\frac{G+1}{2}C^2k^2$\\
\hline
\textbf{2D SDB} & $\frac{2G-1}{G^2}C^2k^2TF$ & $\frac{2G-1}{G^2}C^2k^2$\\
\hline
\textbf{1D SDB} & $\frac{2G-1}{G^2}C^2k^2L$ & $\frac{2G-1}{G^2}C^2k^2$\\
\hline
\end{tabular}
\end{table}

\subsection{Main-processing}
%In the main-processing stage, features are extracted by capturing the relationships between channels, frequencies, and time. %to improve the accuracy of the mapping. 
The main processing comprises sequentially arranged DeFTAN-II blocks that are designed to capture the relationships between channels, frequencies, and time. Each DeFTAN-II block is a combination of F- and T-transformers that employ subgroup processing on spectral and temporal components. 

\begin{figure*}[!t]
\centering
\includegraphics[width=0.8\textwidth]{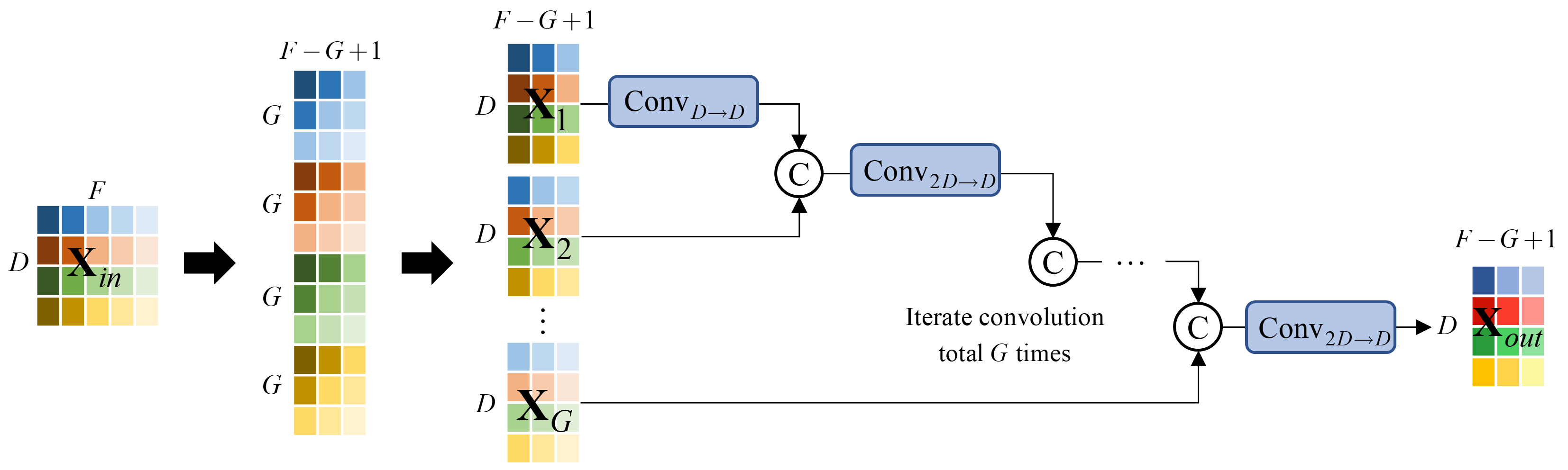}
\caption{Detailed diagram of the 1D split dense block (SDB) of F-transformer}
\label{fig:1d_sdb}
\end{figure*}

\subsubsection{1D Split Dense Block (SDB)}
The first part of the F- and T-transformers employs 1D SDBs that develop features through sequential referencing to shifted input signals. The 1D SDB shares a similar structure with the 2D SDB, but it generates subgroups from shifted input features to enhance the exchange of information with adjacent time or frequency bins. %This idea stems from the conventional adaptive filters that develop filters from the observed signal and apply filters to the unprocessed signal. 
Fig.\,\ref{fig:1d_sdb} provides a detailed diagram of the 1D SDB used in the F-transformer. Initially, the time dimension $T$ of the input feature is permuted into a batch dimension for the F-transformer, resulting in the feature size, excluding the batch dimension, becoming $\mathbf{X}_{in} \in \mathbb{R}^{D \times F}$. Next, the feature undergoes an unfold operation \cite{wang2023tf} with a kernel size of $G$ and a stride of $1$ along the frequency dimension. This unfold operation extracts $G$ subfeatures $\hat{x}_{f} \in \mathbb{R}^{1 \times F-G+1}$ from shifted frequency bins ($[f:F-G+f]$, for $f=1,\cdots,G$) for each channel and stacks them along the channel dimension, resulting in a feature matrix with an increased channel dimension by a factor of $G$. Then, channel shuffle is applied to group subfeatures starting from the same frequency bins. This operation yields $G$ subfeatures $\mathbf{X}_g \in \mathbb{R}^{D \times F-G+1}$ for $g=1,\cdots,G$. 
The remaining process is the same as the 2D SDB, except that 1D Conv is employed instead of 2D Conv. The concatenation and 1D Conv sequentially build up new features by merging information from different frequency bins. %The proposed 1D SDB process is effectively equivalent to increasing the kernel size from $k$ to $k+G-1$ when using a 1D Conv with a kernel size of $k$. Hence, it plays a role in widening the receptive field, leading to enhanced performance.

%In the F-transformer, the input $\mathbf{X}_{\mathrm{in}}$ with $D \times T \times F$ dimensions undergoes an unfold operation [**] with a kernel size of $G$ and stride $1$ along the frequency dimension. Performing unfold pushes the frequency dimension by $G-1$ bin, increasing the channel dimension by a factor of $G$. Therefore, $G$ frequency bins are aligned to the same frequency index and can refer to adjacent frequency bins. This unfolding in frequency dimension yields the tensor size of $GD \times T \times (F-G+1)$. After the time and frequency dimensions are permuted into batch dimensions for F- and T-transformer respectively, the channels of the tensor are rearranged by the channel shuffle, which results in the groups $\mathbf{X}_g = \mathbf{X}_{\mathrm{in}}[:, (g-1)D:gD, :]$. The 1D SDB sequentially concatenates the raw input $\mathbf{X}_g$ and the output from the previous layer, which is filtered again by the convolution layer. The convolution layer on the unfolded feature has the effect of increasing the receptive field for adjacent bins of frequency or time. Therefore, the 1D SDB can develop and apply filters in frequency and channel dimensions with regard to the filtered and raw signals.
%Subsequently, the unfolded input is amplified to $GD \times T \times F$ dimensions by $G$ times and then compressed back to $D$ channels through a 1D SDB. 

\subsubsection{Convolutional Efficient Attention (CEA)}

The developed features are then fed to convolutional efficient attention (CEA). Basically, CEA follows the structure of efficient attention (EA) \cite{shen2021efficient} that produces attention maps from the product of values and keys to reduce the attention map size. However, in CEA, queries and keys are derived by $3\times 3$ convolution layers and gated linear units (GLUs) \cite{dauphin2017language} that extract and gate local features, while values are obtained by linear projections ($1 \times 1$ Conv). This asymmetric processing for values and keys differentiates CEA from the conventional self-attention mechanism \cite{li2019enhancing} using only linear projections or CSA utilizing only convolutions to build queries, keys, and values. In CEA, the keys with emphasized local features are transposed and multiplied to values obtained only by linear projection ($1\times 1$ Conv). Through the cross-attention between keys and values, CEA finds the relations between original features and gated locally emphasized features. 
This attention map is then multiplied with queries to produce an attended output of size $L \times D$, where $L$ indicates the length of sequence dimension ($F-G+1$ or $T-G+1$). The output is then fed into another linear projection layer and added to the input of CEA.
%In the CEA, the dimensions of the $D \times T \times F$-sized tensor are permuted such that the channel dimension and time dimension are mapped to the transformer's embedding dimension and batch dimension, respectively. 
%the attention and FFW of the conventional transformer. By incorporating CNNs into CEA and HCFN, we attempt to consider local spectral relationships for both analysis and synthesis, respectively.  
%the difficulty of capturing local relationships, which is a limitation of transformers. 
%The detailed structures and explanations of CEA and HCFN are as follows.
% When extracting an attention map, CSA uses query and key to utilize the relationships between features with local relationships emphasized. However, CEA uses key and value to extract the attention map and utilizes the relationship between features with emphasized local relationships and features without. This provides more granularity because the difference between the two features is represented in the attention map. In the experiment section, we demonstrate through attention map analysis that simultaneously utilizing local-global relationships like the proposed method can help improve performance. 
%For attention, the norm and add method is employed.  Next, the extracted queries, keys, and values are used to compute the attention output using an efficient attention mechanism [**].  
The CEA can be expressed as follows:
\begin{align}
\begin{aligned}
\mathbf{Q}&=\mathrm{W}_q(\mathrm{GLU}(\mathrm{W}_{c}(\mathbf{X}_{in}))),\\
\mathbf{K}&=\mathrm{W}_k(\mathrm{GLU}(\mathrm{W}_{c}(\mathbf{X}_{in}))),\\
\mathbf{V}&=\mathrm{W}_v(\mathbf{X}_{in}),\\
\mathbf{X}_{out}&=\mathrm{R_{o}}\left(\mathrm{W}_{o}\left(\rho_{q}(\mathbf{Q})\times \mathrm{R_{a}}(\rho_{k}(\mathbf{K}^T\mathbf{V}))/\sqrt{D} \right) \right),
\end{aligned}
\end{align}
where $\mathrm{W}_c$ denotes 1D Conv with kernel size $3$ applied to the input. $\mathrm{W}_q$, $\mathrm{W}_k$, $\mathrm{W}_v$, and $\mathrm{W}_o$ denote pointwise convolutions to obtain the query, key, value, and output, respectively. Dropouts for the attention map and output are indicated as $\mathrm{R}_a$ and $\mathrm{R}_o$, while Softmax functions for queries and keys are denoted as $\rho_{q}$, $\rho_{k}$, respectively. We use multi-head self-attention for CEA with head $h=4$. Table\,\ref{tab:AttCost} shows the complexity of vanilla attention, EA, and CEA for the single-head case, which indicates that CEA is advantageous in reducing computational complexity and memory usage. While CEA exhibits slightly higher computational complexity and memory usage compared to those of EA due to the Conv-GLU structure, computational complexity remains lower than that of vanilla attention, mainly because $L > 4D$. %In Table\,\ref{tab:AttCost}, we consider scaled-dot attention for simplicity.

\subsubsection{Dual-path feedforward network (DPFN)}

The FFW of the proposed model, DPFN, inherits the advantage of the conformer \cite{gulati2020conformer} and CMT \cite{guo2022cmt}, which focuses on capturing local relationships by incorporating convolution into the FFW. Some previous conformer-based models \cite{koizumi2021df, lee2023deft} utilize depthwise convolution to reduce the computational complexity. However, depthwise convolution does not effectively capture relationships across the channel dimension, thus not being the most optimal approach in terms of performance \cite{tan2021efficientnetv2}. In DPFN, one-half of the features (first subgroup) go through dilated convolution (D-conv) to extract features using the increased receptive field, while the rest of the features (second subgroup) are directly concatenated to the output of dilated convolution. The concatenated features are then projected back to the original feature size using linear projection. Thus, unlike previous models \cite{koizumi2021df, lee2023deft}, DPFN applies convolution to only half of the features to mitigate the computational burden. We compared the proposed DPFN to a depthwise convolutional feedforward network (DCFN) using depthwise dilated convolution for the entire features, and to a convolutional feedforward network (CFN) using dilated convolution for the entire feature.
The dilation ratio of the dilated convolution is doubled at every DeFTAN-II block. Finally, the extracted feature is added to the input of the DPFN to produce the output. The formula for DPFN can be written as follows:
\begin{align}
\begin{aligned}
\mathbf{X}_1 &= \mathrm{R}_1(\mathrm{GELU}(\mathrm{W}_{1}(\mathbf{X}_{in})))\\
\mathbf{X}_2 &= \mathrm{PReLU}(\mathrm{LN}(\mathrm{W}_{d}(\mathrm{R}_2(\mathrm{GELU}(\mathrm{W}_{2}(\mathbf{X}_{in}) )))))\\
\mathbf{X}_{out}&=\mathrm{R}_{o}(\mathrm{W}_{o}(\mathrm{Concat}([\mathbf{X}_1, \mathbf{X}_2]))),
\end{aligned}
\end{align}
where $\mathrm{W}_1$, $\mathrm{W}_2$, and $\mathrm{W}_o$ denote the pointwise convolutions applied to the first and second subgroups and FFW output, respectively. $\mathrm{W}_d$ is the dilated convolution for the first half of features. $\mathrm{R}_1$, $\mathrm{R}_2$, and $\mathrm{R}_o$ indicate dropouts for the first and second half of features and the output, respectively.

In Table\,\ref{tab:FFWcost}, the computational complexity and memory usage of DPFN is compared to those of vanilla FFW, CFN, and DCFN. Here, $l$ refers to the kernel size used in the convolution of the FFW. Both metrics of DPFN lie intermediate to those of CFN and DCFN. Nevertheless, as elaborated in Section IV, DPFN demonstrates performance akin to CFN, highlighting that it achieves a favorable trade-off between complexity and performance.

The output of DPFN ($\mathbf{X}_{out} \in \mathbb{R}^{D\times L}$) is then restored to its input size ($D\times F$ or $D \times T$) through a transposed convolution with a kernel size of $G$ and stride $1$. The transposed convolution restores the length of the sequence dimension that was previously reduced by the unfold operation. It is then summed with the transformer input to produce the final output. In the T-transformer, the only difference from the F-transformer is that the time dimension is allocated as the sequence dimension. After $N_b$ iterations of the DeFTAN-II block, the process proceeds to the decoding stage. 

\begin{table}[!t]
\caption{Comparison of Complexity with Various Attentions \label{tab:AttCost}}
\renewcommand{\arraystretch}{1.5}
\centering
\begin{tabular}{ccc}
\hline
& Computational complexity & Memory usage \\
\hline\hline
Vanilla attention & $2DL^2+3D^2L$ & $L^2+3D^2$\\
\hline
EA & $5D^2L$ & $4D^2$\\
\hline
\textbf{CEA} & $(5+2k^2)D^2L$ & $(4+2k^2)D^2$\\
\hline
\end{tabular}
\end{table}

\begin{table}
\caption{Comparison of Complexity with Various FFWs \label{tab:FFWcost}}
\renewcommand{\arraystretch}{1.5}
\centering
\begin{tabular}{ccc}
\hline
& Computational complexity & Memory usage \\
\hline\hline
Vanilla FFW & $8D^2L$ & $8D^2$\\
\hline
DCFN & $8(1+\frac{l^2}{2D})D^2L$ & $8(1+\frac{l^2}{2D^2})D^2$\\
\hline
\textbf{DPFN} & $8(1+\frac{l^2}{2})D^2L$ & $8(1+\frac{l^2}{2})D^2$\\
\hline
CFN & $8(1+2l^2)D^2L$ & $8(1+2l^2)D^2$\\
\hline
\end{tabular}
\end{table}

\subsection{Decoding}
The features extracted by a series of $N_b$ DeFTAN-II blocks are then combined to reconstruct the time-domain waveform of a clean speech. In this stage, a $(3\times 3)$ transposed convolution, denoted as Down-Conv, is utilized to reduce the channel dimension to $2G$, followed by compression using a 2D SDB to further reduce it to $2$. Similar to the 2D SDB structure after Up-Conv, this architecture aggregates the channel and local TF information to reconstruct RI components of the complex spectrogram corresponding to a clean speech. Finally, the estimated complex spectrogram is restored into the time-domain waveform using inverse STFT (iSTFT).

\subsection{Loss functions}
The proposed network is trained by the phase-constrained magnitude (PCM) loss \cite{pandey2021dense}. The PCM loss is defined as the sum of the RI magnitude losses for speech and noise. 

\begin{equation}
L_{P C M}=\frac{1}{2} L_{S M}(\mathbf{S}, \hat{\mathbf{S}})+\frac{1}{2} L_{S M}(\mathbf{N}, \hat{\mathbf{N}}),
\end{equation}
where $\mathbf{S}$, $\hat{\mathbf{S}}$, $\mathbf{N}$, and $\hat{\mathbf{N}}$ denote the spectrogram of the clean speech, estimated speech, original noise, and estimated noise, respectively. The STFT magnitude loss $L_{S M}$ is given as follows.
\begin{equation}
\begin{footnotesize}
\begin{aligned}
L_{S M}(\mathbf{S}, \hat{\mathbf{S}})=\frac{1}{TF} \sum_{t=1}^T & \sum_{f=1}^F \bigg|\left(\Big|S_{\text {real }}(t, f)\Big|-\Big|\hat{S}_{\text {real}}(t, f)\Big|\right) \\
& +\left(\Big|S_{\text {imag }}(t, f)\Big|-\Big|\hat{S}_{\text {imag }}(t, f)\Big|\right)\bigg|,
\end{aligned}
\end{footnotesize}
\end{equation}
where $S_{\text {real }}(t, f)$, $S_{\text {imag }}(t, f)$ and $\hat{S}_{\text {real }}(t, f)$, $\hat{S}_{\text {imag }}(t, f)$ correspond to the $(t, f)$-th RI components of $\mathbf{S}$ and $\hat{\mathbf{S}}$, respectively.

\section{Experiments}
\subsection{Datasets}
Three distinct datasets, spatialized WSJCAM0, spatialized DNS challenge, and L3DAS22, were utilized for the training and testing of the proposed model. Speech data were spatialized through the convolution with multichannel room impulse responses (RIRs) simulated in diverse acoustic environments to avoid bias of the model towards a particular environment. Then, the model was tested in a real-world environment to demonstrate its real-world performance.
%Each simulation corresponds to a different dataset with its specific environment. This approach demonstrates that the proposed model is not biased towards a particular dataset and can achieve high performance in diverse environments. 
%To secure diversity in the acoustic environment, each simulation was conducted 
%Furthermore, we plan to showcase experiments and results in real-world environments in a later section, using the model trained on these datasets.
\subsubsection{Spatialized WSJCAM0 dataset}
First, we constructed a highly reverberant dataset using the WSJCAM0 corpus \cite{robinson1995wsjcamo} sampled at 16 kHz. We simulated spatialized speech signals using a 4-channel circular microphone array with a radius of 10\,cm. The cuboid rooms were generated by randomly sampling the width, depth, and height dimensions from uniform distributions within the ranges [5, 10] m, [5, 10] m, and [3, 4] m, respectively. The RIRs were generated using the image source method and \textit{pyroomacoustics} library \cite{scheibler2018pyroomacoustics}, and the reverberation time (T60) of each room was varied between [0.2, 1.3] s. Individual speech data were convolved with different RIRs, so none of the spatialized data shared the identical room environment. Additionally, the room environments for the training, validation, and test sets were all different. For the training dataset, speech utterances of 4-second duration were randomly selected from the corpus. Spatialized speeches were then merged with noise signals obtained from the 1st, 3rd, 5th, and 7th channel noises in the REVERB challenge dataset (weak air-conditioner noises) \cite{kinoshita2016summary}. 
The signal-to-noise ratio (SNR) between the speech and noise signals at the reference microphone (channel 1) was set between [5, 25]\,dB. This dataset includes stationary and relatively weak noises, with a greater emphasis on reverberation than noise. The algorithm used to create the dataset is described in \cite{wang2020multi}.

\subsubsection{Spatialized DNS challenge dataset}
Next, we synthesized another noisy-reverberant dataset using the DNS Challenge 2020 corpus \cite{reddy2020interspeech} for both the speech and noise components sampled at 16 kHz. The same circular microphone array and room configuration were used. In this dataset, SNRs range between [-10, 10]\,dB, and the T60 range was from [0.2, 1.2]\,s. The noise data in this dataset is non-stationary, and the levels are higher than that in the WSJCAM0 dataset, making noise removal more challenging. Since the original DNS challenge dataset contains single-channel data, we spatialized both speeches and noises using a similar procedure as in the spatialized WSJCAM0 dataset described in \cite{pandey2022tparn}.

\subsubsection{L3DAS22 Challenge dataset}
The last dataset used for evaluation is the L3DAS22 challenge \cite{guizzo2022l3das22} dataset proposed as part of the recent ICASSP 2022 challenges. This dataset includes speech recordings simulated in 3D office environments with varying speaker positions. In the L3DAS22 dataset, noisy reverberant speeches were simulated for two 4-channel Ambisonic microphone arrays, using speech and noise signals from LibriSpeech \cite{panayotov2015librispeech} and FSD50k \cite{fonseca2021fsd50k}, respectively. 
The SNR ranges from 6 to 16\,dBFS (referring to the signals' RMS amplitude). One microphone array (mic A) was fixed at the center of the room, while the other array (mic B) was positioned 20\,cm away from the center. The room configuration, measuring 6\,m in width, 3\,m in depth, and 5\,m in height, remained the same for both train and test. The source positions were uniformly and randomly sampled to ensure that each position was unique. The data was also sampled at a rate of 16 kHz.

\subsection{Parameter setup}
For the STFT, a Hamming window of 32\,ms length and hop size of 16\,ms (50\% overlap) were used. Since all three datasets mentioned above sampled at a rate of 16 kHz, each STFT frame comprised 512 samples. %, and the RI components of complex spectra are utilized. 
The number of DeFTAN-II blocks was set to 6 for the base model and 12 for the large model. The parameters used in this work are listed in Table \ref{tab:hyperparameters}. 

\begin{table}[!t]
\caption{Hyperparameters of the proposed model\label{tab:hyperparameters}}
\renewcommand{\arraystretch}{1.5}
\centering
\begin{tabular}{c|c|c}
\hline
Symbol & Description & Value \\
\hline\hline
$C$ & Number of output channels of Up Conv & 256 \\
\hline
$G$ & Number of groups of SDB (Both 2D and 1D) & 4 \\
\hline
$D$ & Number of input channels of DeFTAN-II blocks & 64 \\
\hline
$I$ & Kernel size of unfold and transposed convolution & 4 \\
\hline
$J$ & stride of unfold and transposed convolution & 1 \\
\hline
$N_b$ & Number of DeFTAN-II blocks & 6 \\
\hline
$k$ & Kernel size of convolution used in SDB and CEA & 3 \\
\hline
$h$ & head of multi-head self-attention used in CEA & 4 \\
\hline
$l$ & Kernel size of convolution used in DPFN & 5 \\
\hline
\end{tabular}
\end{table}

The Adam optimizer\cite{kingma2014adam} with an initial learning rate of $4\times 10^{-4}$ was used for training, and the learning rate was halved if the validation loss did not decrease for five consecutive epochs. The model was trained for 100 epochs, and the model from the epoch with the lowest validation loss was selected as the final model. The batch size was set to 1. The training was performed on eight GeForce RTX 3090 GPUs. The speech enhancement performance of the models trained on each dataset was evaluated using three objective measures: scale-invariant signal-to-distortion ratio (SI-SDR) \cite{le2019sdr}, perceptual assessment of speech quality (PESQ) \cite{rix2001perceptual}, and short-time objective intelligibility (STOI) \cite{taal2010short}. Additionally, the parameter sizes and computational complexities (MAC/s) were also compared.

\subsection{Experiment procedure}
First, we carried out the parameter study to evaluate the efficacy of the proposed blocks comprising DeFTAN-II. We ablated or replaced the attention, FFW blocks, and SDBs, and investigated the performance change on the spatialized WSJCAM0 dataset. Next, we compared the proposed model with state-of-the-art models \cite{pandey2022tparn, pandey2022multichannel, lee2023deft, wang2023tf, lu2022towards, zhang2022multi, li2022pcg}. The comparison was made by testing corresponding models on the spatialized WSJCAM0, DNS challenge, and L3DAS22 challenge datasets. 

%The comparison with TF GridNet and parameter study of the proposed blocks were conducted by training and evaluating models using the spatialized WSJCAM0 dataset. 

%For each experiment, we trained the model using the train set of each dataset and compared the results obtained on the test set.

\section{Results}
\subsection{Parameter study}

\begin{table*}[!t]
\caption{Parameter study results\label{tab:param}}
\renewcommand{\arraystretch}{1.5}
\centering
\begin{tabular}{c||c|c|c|c|c|c|c|c|c||c|c|c|c|c}
\hline
Model & $C$ & $G$ & $D$ & $I$ & $J$ & 2D SDB & 1D SDB & Attention & FFW & SI-SDR & PESQ & STOI & Param. & MAC/s \\
\hline\hline
1 & 64 & - & 64 & 1 & 1 & - & - & EA & FFW & 16.0 & 3.75 & 0.980 & \textbf{1.2 M} & \textbf{20.4 G}\\
\hline \hline
2 & 64 & - & 64 & 1 & 1 & - & Dense & EA & FFW & 16.5 & 3.80 & 0.980 & 3.5 M & 56.4 G\\
\hline
3 & 64 & - & 64 & 1 & 1 & - & Dense & CSA & FFW & 16.4 & 3.80 & 0.979 & 4.3 M & 70.6 G\\
\hline
\textbf{4} & 64 & - & 64 & 1 & 1 & - & Dense & \textbf{CEA} & FFW & 17.1 & 3.84 & 0.981 & 3.8 M & 61.2 G\\
\hline \hline
5 & 64 & - & 64 & 1 & 1 & - & Dense & EA & DCFN & 16.7 & 3.81 & 0.980 & 3.7 M & 56.9 G \\
\hline 
\textbf{6} & 64 & - & 64 & 1 & 1 & - & Dense & EA & \textbf{DPFN} & 17.3 & 3.86 & 0.981 & 4.5 M & 72.5 G\\
\hline
7 & 64 & - & 64 & 1 & 1 & - & Dense & EA & CFN & 17.5 & 3.85 & 0.982 & 7.4 M & 121.4 G\\
\hline  \hline
8 & 64 & - & 64 & 1 & 1 & - & Dense & \textbf{CEA} & \textbf{DPFN} & 17.7 & 3.89 & 0.983 & 4.8 M & 77.3 G\\
\hline
% \textbf{9} & 64 & 4 & 64 & 4 & 1 & - & \textbf{SDB} & EA & FFW & 17.5 & 3.84 & 0.984 & 2.4 M & 39.2 G\\
% \hline
% \textbf{10} & 256 & 4 & 64 & 4 & 1 & \textbf{SDB} & \textbf{SDB} & EA & FFW & 17.8 & 3.86 & 0.984 & 2.7 M & 43.8 G\\
% \hline
9 & 64 & 4 & 64 & 4 & 1 & - & \textbf{SDB} & \textbf{CEA} & \textbf{DPFN} & 18.1 & 3.91 & 0.986 & 3.7 M & 59.9 G\\
\hline
\textbf{10} & 256 & 4 & 64 & 4 & 1 & \textbf{SDB} & \textbf{SDB} & \textbf{CEA} & \textbf{DPFN} & \textbf{18.4} & \textbf{3.92} & \textbf{0.987} & 4.0 M & 64.5 G\\
\hline
\end{tabular}
\end{table*}

We organized the parameter study as follows: the baseline model (Model\,1) was the model with F- and T-transformers including only the EA and vanilla FFW networks. Model\,2 used vanilla dense blocks in the F- and T-transformers, and Model\,3 employed the conventional CSA as an attention mechanism instead of EA, which was compared with Model\,4 utilizing the proposed CEA. Models 5 -- 7 were designed to examine the impact of using convolution and DPFN in the FFW. In Models 5, 6, and 7, we replaced the vanilla FFW structure of Model\,2 with DCFN, DPFN, and CFN, respectively. Model\,8 included the best attention and FFW structures (CEA and DPFN) determined from the previous comparisons. Based on this model, we tested the contribution of SDBs in Models 9 and 10 by substituting the vanilla dense blocks with SDBs. 

The results of the parameter studies are shown in Table\,\ref{tab:param}. First, we can check the efficacy of the 1D SDB by comparing Models 1 and 2. The 1D SDB allows for capturing relationships in the channel dimension, resulting in enhanced performance (+0.5\,dB in SI-SDR), but it also significantly increases the computational complexity (from 20.4\,G to 56.4\,G). 

The following experiment compares the attention mechanisms through models using EA, CSA, and CEA. By comparing Models 2, 3, and 4, we can see that the CEA employing the Conv-GLU structure exhibits a slight increase in computational complexity (from 56.4\,G to 61.2\,G) and parameter size (from 3.47\,M to 3.77\,M), but greatly improves performance (+0.6\,dB in SI-SDR). In contrast, Model\,3 using CSA performs worse (-0.1\,dB in SI-SDR) despite its larger parameter size (4.3\,M) and higher computational complexity (70.6\,G) than Model\,4 using CEA. This suggests that CEA employing subgroup processing outperforms other attention mechanisms.

In the next experiment, we compared the model using DPFN (Model\,6) with the models using various FFWs. The comparison of Model\,6 to Model\,2 demonstrates a significant increase in computational complexity (from 56.4\,G to 72.5\,G) due to the use of dilated convolution. However, the performance also highly increases (+0.8\,dB in SI-SDR) due to the ability to capture channel and local TF relationships. 
In the case of DCFN that uses depthwise dilated convolution (Model\,5), the performance improvement from vanilla FFW (Model\,2) is relatively small (+0.2\,dB in SI-SDR). Although the parameter size and computational complexity of dilated convolution are higher than those of depthwise dilated convolution, DPFN achieves performance similar to CFN (Model\,7) with much lower computational complexity and parameter size. 
We also experimented with a model that combines CEA and DPFN (Model\,8). The comparison of Models 6 and 8 stresses that we can obtain an additional performance improvement (16.5\,dB to 17.5\,dB over SI-SDR) from the combination of CEA and DPFN.
% The following experiment compares the model using HCFN and the model using FFW. By comparing Model\,2 and Model\,6, it can be observed that the computational requirements significantly increase (from 56.4 G to 72.5 G) due to the use of standard convolution, which requires capturing many channel relationships. However, the performance also increases (from 16.5\,dB to 17.3\,dB in terms of SI-SDR) because of the ability to capture channel and local TF relationships. Furthermore, as the index of the DRACHEN block increases, the dilation also increases, allowing for a wider receptive field. Additionally, comparing Model\,2 with Model\,4, and Model\,2 with Model\,5 allows for comparing the model using HCFN with the one using CFN. If the model uses CFN instead of HCFN, the performance change is not significant, but the parameter size (from 4.46 M to 7.43 M) and computational requirements (from 72.5 G to 121.4 G) increase more. Therefore, in the proposed model, we used HCFN, which has lower computational requirements and higher performance compared to those of CFN. We also conducted experiments on models that combine CEA and HCFN. By comparing Model\,2 with Model\,3, Model\,4, and Model\,6, it can be observed that when CEA and HCFN are used together, there is additional performance improvement (from 16.5\,dB to 17.5\,dB in terms of SI-SDR) compared to the performances of using each model separately.

In Models 8 and 9, we conducted ablation studies on the dense block and SDB. Model\,9 using SDB is significantly lighter (59.9\,G) in computational complexity than Model\,9 using dense block (77.3\,G) but achieves higher performance (+0.4\,dB in SI-SDR). This result demonstrates that the feature extraction from the concatenation of unfolded data and channel reduction using 1D Conv can efficiently capture crucial features with low complexity.  
% While the dense block is known to be useful for capturing relationships between channels, it has limitations in terms of the maximum number of available channels when using the same memory, as it simply stacks the previous channels and uses convolution. On the other hand, SDB can extract important features by using more channels than the original number of channels and allowing for a larger maximum number of available channels compared with the case of using a dense block.
%There is a significant difference in the way SDB concatenates grouped sub-features rather than simply concatenating previous features used in convolution in the dense block. 
%Therefore, extracting features in the channel dimension with $G$ times more features and then shrinking them by $G$ times with SDB provides better performance improvement in speech enhancement. 

In total, we obtained a +1.6\,dB increase in SI-SDR by incorporating the proposed 1D SDB, CEA, and DPFN architectures as compared to Model\,2 using the plain dense block, efficient attention, and FFW blocks. Lastly, by utilizing 2D SDB in the encoder and decoder (Model\,10), SI-SDR was further improved to 18.4\,dB at the expense of 4.6\,G increase in computational complexity. 

From the parameter studies conducted, it is evident that all the proposed blocks in DeFTAN-II significantly contribute to its final performance. Notably, the proposed blocks share a common characteristic; they involve information exchange between two subgroups of features that are processed and unprocessed by local convolution. %Specifically, SDB sequentially concatenates these two subgroups to refine features, and CEA utilizes cross-attention between the two subgroups. DPFN also generates and concatenates two subgroups to reduce computational complexity while minimizing the loss in enhancement performance. 
In the subsequent experiment, we demonstrate that the proposed model with this subgroup processing outperforms state-of-the-art multichannel speech enhancement models in terms of both speech enhancement performance and computational complexity. 

%using CEA and DPFN in the transformer structure to simultaneously capture local and global information, and applying SDB before and after each block (Up and Down Conv, F/T-transformer) for aggregate channel and local TF relations, is effective in extracting features and improving performance.

% \input{Attention_analysis}

% \begin{figure}
% \centering
% \includegraphics[width=3.5in]{Figure/Fig_att_map.png}
% \caption{Attention map of T-transformer using (a) proposed convolutional efficient attention (CEA) (b) convolutional self-attention (CSA)}
%\caption{Attention map of T-transformer with using (a) Convolutional Self-Attention (CSA) (b) Proposed Conv-Efficient Attention (CEA)}
% \label{fig:att_map}
% \end{figure}

\subsection{Comparison to multichannel speech enhancement models}
The models we considered for comparison are TPARN \cite{pandey2022tparn}, ADCN \cite{pandey2022multichannel}, DeFTAN \cite{lee2023deft}, and TF GridNet \cite{wang2022tf}. These models have exhibited state-of-the-art performance utilizing transformer architectures. Specifically, DeFTAN is our previous model based on the dense, F-transformer, and T-conformer, which is the most similar to the proposed model in terms of the model architecture. TF GridNet also utilizes the unfold-transposed convolution processing, and its parameters were set to the values reported as the best model \cite{wang2022tf}: 75\% overlap, 64 channel dimensions, and the same unfold-transposed convolution setting ($I=4$, $J=1$) as the proposed model. 
We also compared the performance of models in 2-stage settings. The 2-stage models of Table\,\ref{tab:compWSJCAM0_and_DNS} represent models using the 2-stage processing methods described in \cite{pandey2022multichannel, lee2023deft, wang2022tf}. 
For comparison with different models, we trained and tested the proposed model on three datasets. Through the comparison using different datasets, we examined the consistency and possible data bias of the models.

\begin{table*}[]
\caption{Performance comparison on Spatialized WSJCAM0 and Spatialized DNS challenge (4CH)\label{tab:compWSJCAM0_and_DNS}}
\renewcommand{\arraystretch}{1.5}
\centering
\begin{tabular}{c||ccc|ccc|c|c}
\hline
\multirow{2}{*}{Model}                                                & \multicolumn{3}{c|}{WSJCAM0}                                                             & \multicolumn{3}{c|}{DNS challenge}                                                       & \multirow{2}{*}{\begin{tabular}[c]{@{}c@{}}Param.\\ size\end{tabular}} & \multirow{2}{*}{MAC/s} \\ \cline{2-7}
                                                                      & \multicolumn{1}{c|}{SI-SDR}        & \multicolumn{1}{c|}{PESQ}          & STOI           & \multicolumn{1}{c|}{SI-SDR}        & \multicolumn{1}{c|}{PESQ}          & STOI           &                                                                        &                        \\ \hline\hline
Unprocessed                                                           & \multicolumn{1}{c|}{-4.0}          & \multicolumn{1}{c|}{1.93}          & 0.707          & \multicolumn{1}{c|}{-7.8}          & \multicolumn{1}{c|}{1.38}          & 0.593          & -                                                                      & -                      \\ \hline
TPARN \cite{pandey2022tparn}                                          & \multicolumn{1}{c|}{10.4}         & \multicolumn{1}{c|}{3.43}          & 0.969          & \multicolumn{1}{c|}{8.4}           & \multicolumn{1}{c|}{2.75}          & 0.919          & 3.2 M                                                                  & 71.8 G                 \\ \hline
ADCN  \cite{pandey2022multichannel}                                   & \multicolumn{1}{c|}{12.0}          & \multicolumn{1}{c|}{3.42}          & 0.973          & \multicolumn{1}{c|}{7.8}           & \multicolumn{1}{c|}{2.84}          & 0.923          & 9.3 M                                                                  & 72.8 G                 \\ \hline
DeFTAN \cite{lee2023deft}                                             & \multicolumn{1}{c|}{15.7}          & \multicolumn{1}{c|}{3.63}          & 0.981          & \multicolumn{1}{c|}{9.9}           & \multicolumn{1}{c|}{3.01}          & 0.924          & \textbf{2.7 M}                                                         & 95.6 G                 \\ \hline
TF GRidNet \cite{wang2022tf}                                          & \multicolumn{1}{c|}{17.5}          & \multicolumn{1}{c|}{3.82}          & 0.983          & \multicolumn{1}{c|}{10.9}          & \multicolumn{1}{c|}{3.18}          & 0.939          & 14.7 M                                                                 & 462.2 G                \\ \hline
\renewcommand{\arraystretch}{1.0}
%\textbf{\begin{tabular}[c]{@{}c@{}}DeFTAN-II\\ (base)\end{tabular}}  
\textbf{DeFTAN-II (base)} & \multicolumn{1}{c|}{\textbf{18.4}} & \multicolumn{1}{c|}{\textbf{3.92}} & \textbf{0.987} & \multicolumn{1}{c|}{\textbf{12.0}} & \multicolumn{1}{c|}{\textbf{3.35}} & \textbf{0.953} & 4.0 M                                                                  & \textbf{64.5 G}        \\ \hline\hline
\renewcommand{\arraystretch}{1.0}
\begin{tabular}[c]{@{}c@{}}2 stage\\ TPARN-ADCN \cite{pandey2022multichannel}\end{tabular}          & \multicolumn{1}{c|}{13.8}          & \multicolumn{1}{c|}{3.64}          & 0.980          & \multicolumn{1}{c|}{10.0}          & \multicolumn{1}{c|}{2.99}          & 0.937          & 40 M                                                                   & 360.0 G                \\ \hline
\renewcommand{\arraystretch}{1.0}
\begin{tabular}[c]{@{}c@{}}2 stage\\ DeFTAN \cite{lee2023deft}\end{tabular}              & \multicolumn{1}{c|}{15.8}          & \multicolumn{1}{c|}{3.73}          & 0.982          & \multicolumn{1}{c|}{10.7}          & \multicolumn{1}{c|}{3.10}          & 0.930          & \textbf{5.4 M}                                                         & 191.2 G                \\ \hline
\renewcommand{\arraystretch}{1.0}
\begin{tabular}[c]{@{}c@{}}2 stage\\ TF GridNet \cite{wang2022tf}\end{tabular}          & \multicolumn{1}{c|}{18.3}          & \multicolumn{1}{c|}{3.90}          & 0.986          & \multicolumn{1}{c|}{11.7}          & \multicolumn{1}{c|}{3.29}          & 0.948          & 29.6 M                                                                 & 924.6 G                \\ \hline
\renewcommand{\arraystretch}{1.0}
%\textbf{\begin{tabular}[c]{@{}c@{}}DeFT-AN-II\\ (large)\end{tabular}} 
\textbf{DeFTAN-II (large)} & \multicolumn{1}{c|}{\textbf{20.6}} & \multicolumn{1}{c|}{\textbf{4.02}} & \textbf{0.991} & \multicolumn{1}{c|}{\textbf{13.0}} & \multicolumn{1}{c|}{\textbf{3.43}} & \textbf{0.959} & 7.7 M                                                                  & \textbf{124.0 G}       \\ \hline
\end{tabular}
\end{table*}

\subsubsection{Spatialized WSJCAM0 dataset}
The results of the experiments on the spatialized WSJCAM0 dataset are presented in Table\,\ref{tab:compWSJCAM0_and_DNS}. The experimental results show that the proposed model exhibits significantly higher performance compared to other multichannel speech enhancement models in highly reverberant environments. The proposed DeFTAN-II achieves a performance of 18.4\,dB in terms of SI-SDR on the test set, and the large DeFTAN-II achieves a performance of 20.6\,dB. These performances are higher than those of the former best model (TF GridNet), and DeFTAN-II (base) accomplishes these only using 27\% of parameter size and 14\% of computational complexity. 

\begin{figure}
\centering
\includegraphics[width=3.7in]{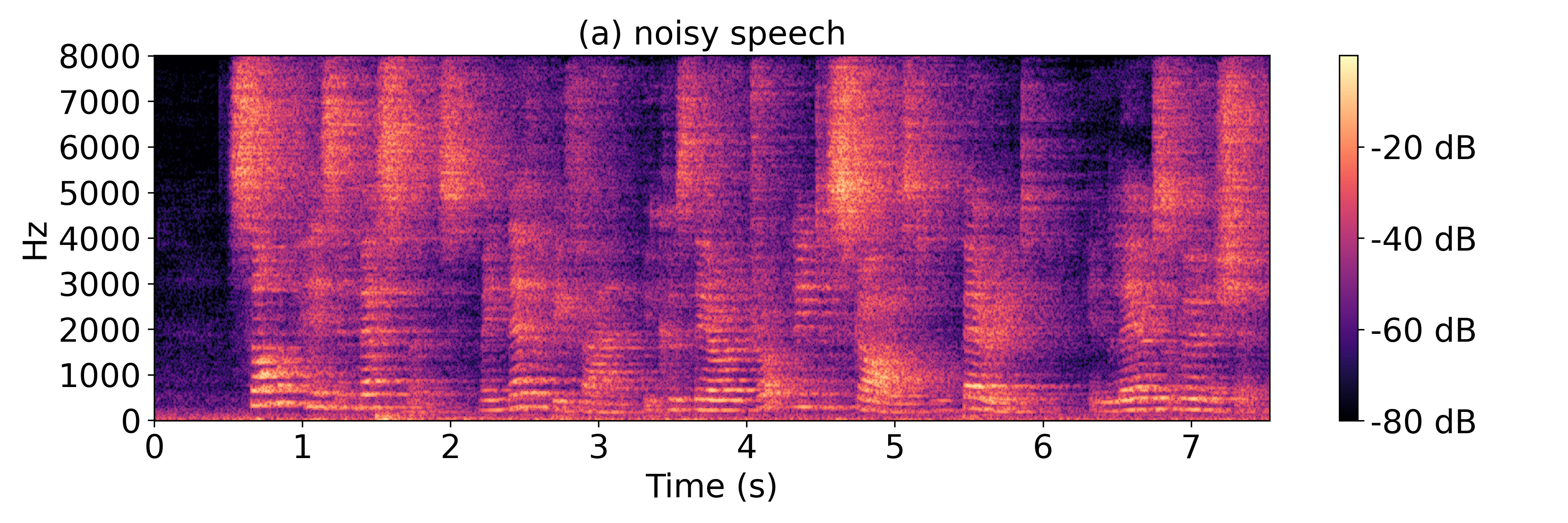}
\includegraphics[width=3.7in]{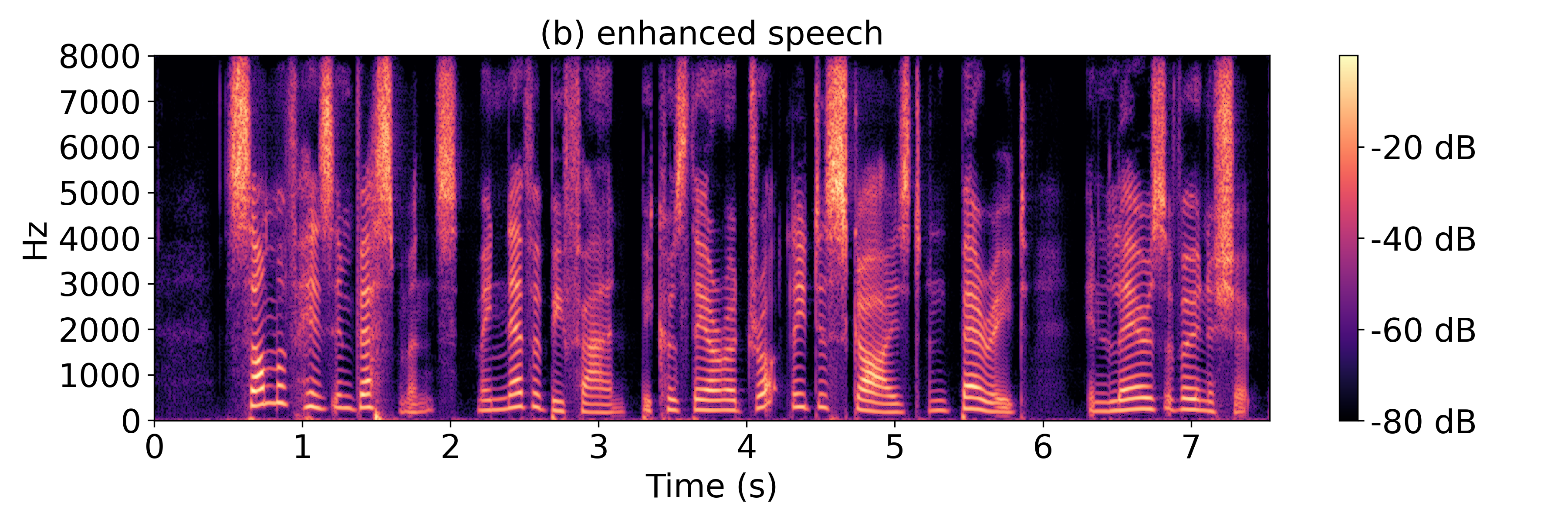}
\includegraphics[width=3.7in]{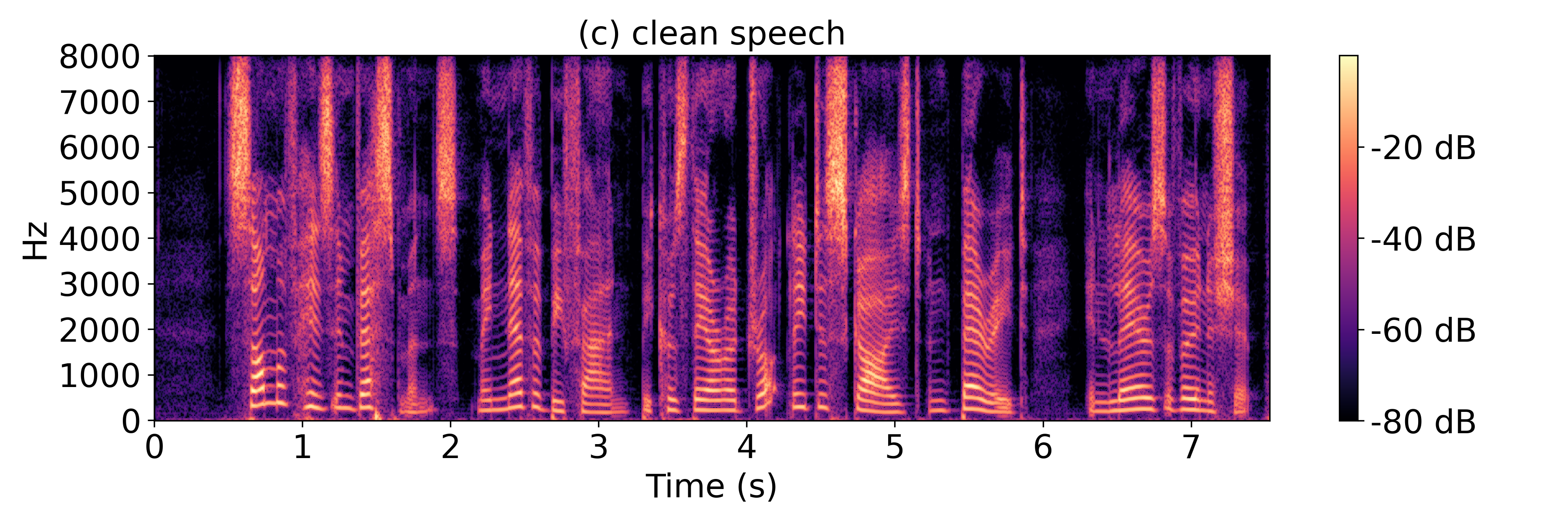}
\caption{Spectrogram example of spatialized WSJCAM0 dataset, (a) noisy reverberant speech, (b) speech enhanced by DeFTAN-II (large), (c) clean speech}
\label{fig:WSJ}
\end{figure}

Fig.\,\ref{fig:WSJ} shows spectrograms of the noisy, enhanced, and clean signals, demonstrating that speech is well restored even in environments with a T60 exceeding 1 second. Particularly, the STOI metric of DeFTAN-II scored 0.987 and 0.991 for the base and large models, respectively, which indicates that the enhanced speech is nearly indistinguishable from the clean speech. 

\subsubsection{Spatialized DNS Challenge dataset}
\begin{figure}
\centering
\includegraphics[width=3.7in]{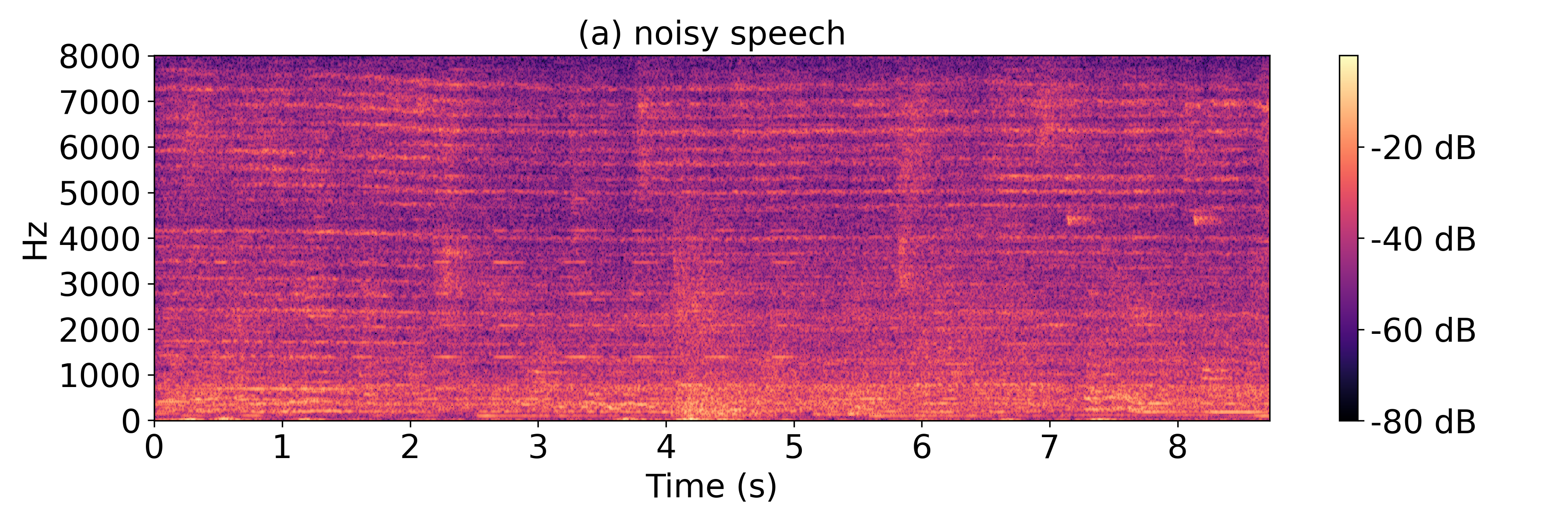}
\includegraphics[width=3.7in]{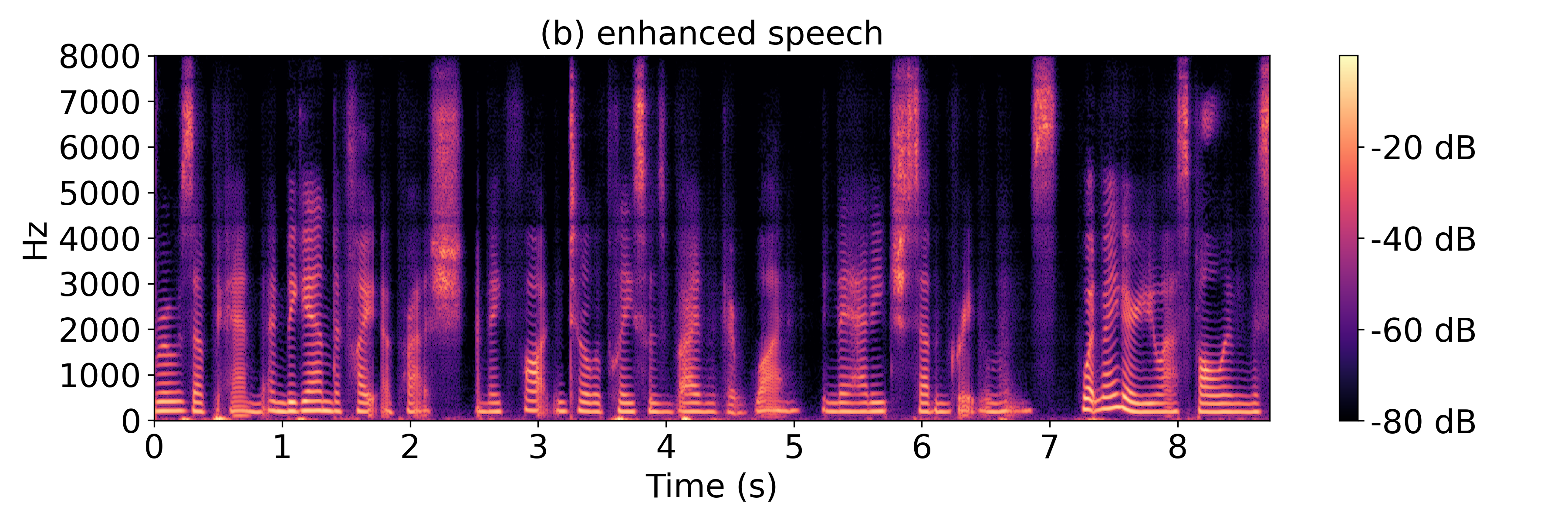}
\includegraphics[width=3.7in]{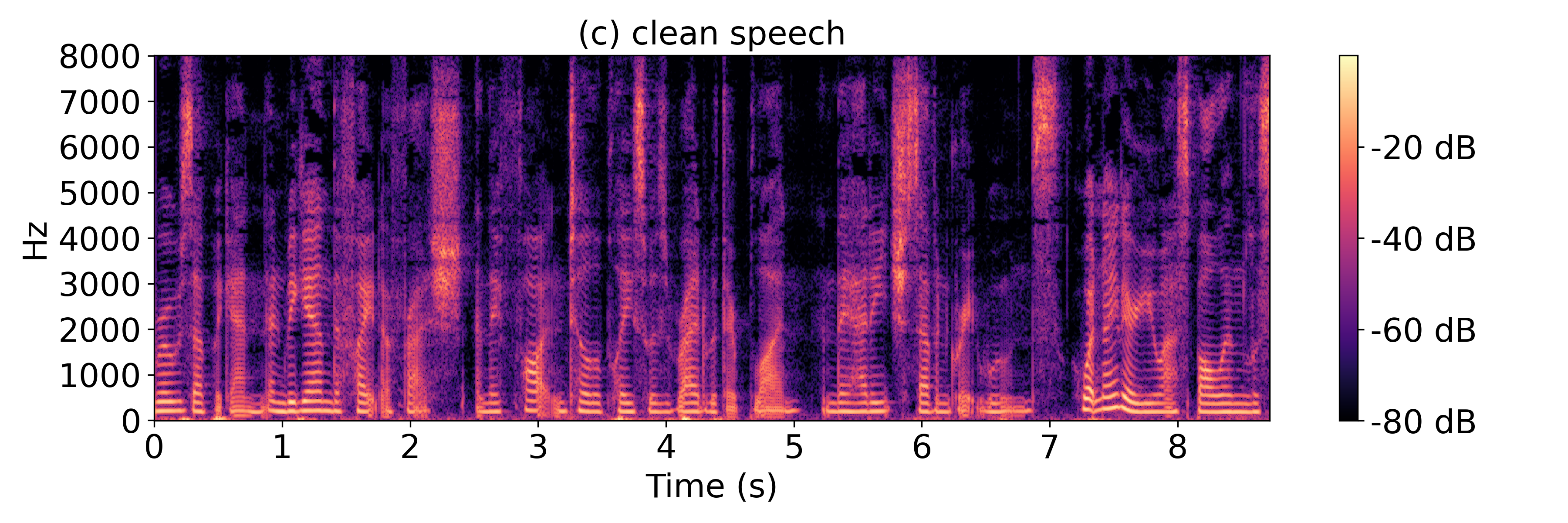}
\caption{Spectrogram example of spatialized DNS challenge dataset, (a) noisy reverberant speech, (b) speech enhanced by DeFTAN-II (large), (c) clean speech}
\label{fig:DNS}
\end{figure}

Table\,\ref{tab:compWSJCAM0_and_DNS} presents the comparison of models trained and tested on the spatialized DNS challenge dataset. 
%This dataset exhibits diverse noise characteristics compared to the previous dataset, and it poses challenges due to non-stationary noise and a low SNR environment, making noise removal relatively difficult. 
The performances of all models are lower than those on the spatialized WSJCAM0 dataset because this dataset includes non-stationary noise and a low SNR environment. However, the results also show that the proposed model (base) outperforms other single- and two-stage models in terms of all evaluation metrics considered. These improvements signify the robustness of DeFTAN-II in challenging noisy environments. Using the large DeFTAN-II further improves all metrics, surpassing other 2-stage models by 2-3\,dB in SI-SDR. The spectrograms of Fig.\,\ref{fig:DNS} present a very low SI-SDR environment near -20 dB, where speech is nearly discernible from the noisy signal. In an environment with such low SNR, DNNs struggle to restore the formants in the high-frequency band, and speech recognition becomes difficult. However, the spectrogram of the speech enhanced by DeFTAN-II shows well-restored high-frequency formants, which also leads to high PESQ scores. 

\subsubsection{L3DAS22 Challenge dataset}
Comparison on the L3DAS22 dataset is presented in Table\,\ref{tab:compL3DAS}. In this experiment, the comparison set was constructed from the top-rank models on L3DAS22 (PCG-AIID \cite{li2022pcg}, BaiduSpeech \cite{zhang2022multi}, ESP-SE \cite{lu2022towards}) and TF GridNet. For DeFTAN-II, we used the SI-SDR loss instead of PCM loss, because it performed better. As the evaluation metrics, we incorporated both the word error rate (WER) and STOI metrics according to the challenge guideline. The speech recognition performance measured by WER was obtained using Wav2Vec2.0 \cite{baevski2020wav2vec} pre-trained on the Librispeech 960h dataset \cite{panayotov2015librispeech}. The total score of each model is given by $\mathrm{Score}=(\mathrm{STOI}+(1-\mathrm{WER}))/2$. Similar to the results with other datasets, the base and large DeFTAN-II models surpass the performance of other models in all metrics. The base DeFTAN-II demonstrates superior performance with over seven times lower computations compared to that in TF GridNet, highlighting its significance for efficient model design.

\begin{table}
\caption{Performance comparison on L3DAS22 Challenge (8CH)\label{tab:compL3DAS}}
\setlength{\tabcolsep}{5pt}
\renewcommand{\arraystretch}{1.5}
\centering
\begin{tabular}{c||c|c|c|c|c}
\hline
% \multicolumn{1}{c||}{Model}       & \multicolumn{1}{c|}{WER (\%)} & \multicolumn{1}{c|}{STOI} & \multicolumn{1}{c|}{Score} & \multicolumn{1}{c|}{\multirow{2}{*}{Param.}} & \multicolumn{1}{c}{\multirow{2}{*}{MAC/s}} \\ 
% \cline{1-4}
% \multicolumn{1}{c||}{Baseline} & \multicolumn{1}{c|}{21.20}   & \multicolumn{1}{c|}{0.878} & \multicolumn{1}{c|}{0.833} & \multicolumn{1}{c|}{} & \multicolumn{1}{c}{} \\
% \hline\hline
Model & WER (\%) & STOI & Score & \renewcommand{\arraystretch}{1.0}\begin{tabular}{@{}c@{}} Param. \\ size \end{tabular} & MAC/s \\
\hline\hline
Unprocessed & 21.20 & 0.878 & 0.833 & - & - \\
\hline
PCG-AIID \cite{li2022pcg} & 3.20 & 0.972 & 0.970 & - & - \\ 
\hline
BaiduSpeech \cite{zhang2022multi} & 2.50 & 0.975 & 0.975 & - & - \\ 
\hline
ESP-SE \cite{lu2022towards} & 1.89 & 0.987 & 0.984 & - & - \\ 
\hline
TF GridNet \cite{wang2022tf} & 1.68 & 0.988 & 0.985 & 14.7 M & 462.2 G \\ 
\hline
\renewcommand{\arraystretch}{1.0}
\begin{tabular}{@{}c@{}}\textbf{DeFTAN-II} \\ \textbf{(base)} \end{tabular} & \textbf{1.64} & \textbf{0.989} & \textbf{0.986} & \textbf{4.0 M} & \textbf{64.5 G} \\
\hline\hline
\renewcommand{\arraystretch}{1.0}
\begin{tabular}{@{}c@{}}2stage \\ TF GridNet \end{tabular} & 1.29 & 0.994 & 0.990 & 29.4 M & 924.6 G \\ 
\hline
\renewcommand{\arraystretch}{1.0}
\begin{tabular}{@{}c@{}}\textbf{DeFTAN-II} \\\textbf{(large)} \end{tabular} & \textbf{1.25} & \textbf{0.994} & \textbf{0.991} & \textbf{7.7 M} & \textbf{124.0 G} \\
\hline
\end{tabular}
\end{table}

\subsection{Real-world Experimental setting}
\begin{figure}
\centering
\includegraphics[width=2.5in]{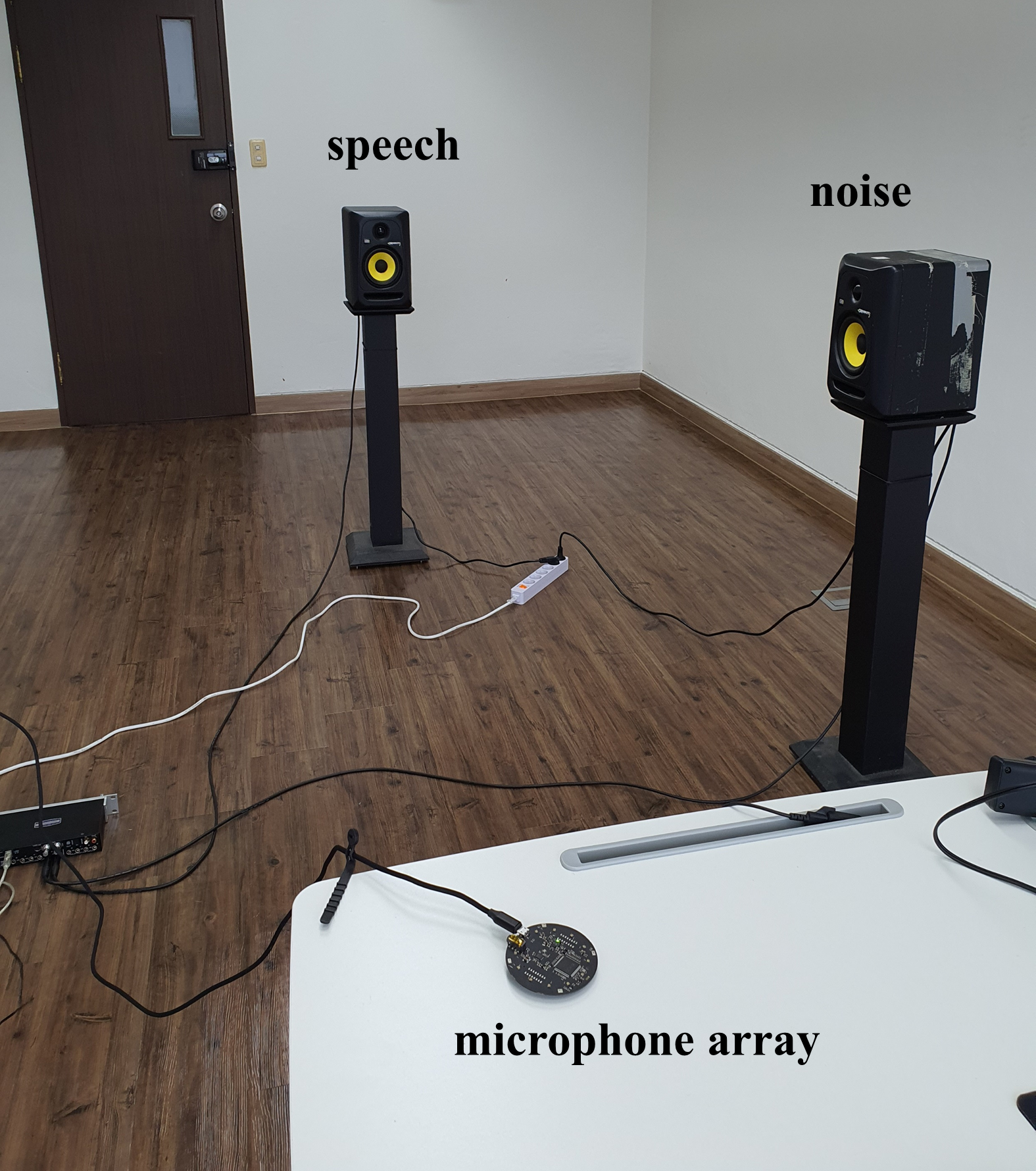}
\caption{Real-world environment experiment setup in the empty office environment.}
\label{fig:realconfig}
\end{figure}

Contemporary multichannel speech enhancement models have primarily focused on improving performance using simulated datasets. However, it is also important to investigate their real-world performance by exposing them to real-world environments. We explored the feasibility of DeFTAN-II in a real environment by directly exposing the pre-trained model to real-world data without fine-tuning. Fig.\,\ref{fig:realconfig} shows the photograph of the real-world experimental setup. We placed two speakers (KRK Rockit 5) reproducing noise and speech signals, as well as a 4-channel circular microphone array (ReSpeaker Mic Array), inside a small office room of 0.5\,s reverberation time (T60). The speakers and microphone were positioned at heights of 1\,m and 0.8\,m above the floor, respectively, with a minimum distance of 50\,cm between the walls and between the speakers and the microphones. The size of the microphone array (2.7\,cm radius) is different from that used for the training (10\,cm radius), so we can examine how the trained model can adapt to this array geometry change. We used data from LibriSpeech to produce speech signals and DNS Challenge 2020 data for noise signals. We used a large DeFTAN-II model trained on a simulated DNS challenge dataset to enhance noisy speech measured in real-world environments. Speech and noise signals fed into the loudspeaker are scaled such that the SNR becomes $5$\,dB in a signal level, and thus, actual SNRs at the microphone array position vary depending on the placement of loudspeakers and microphones. We ran a total of 100 experiments with different speech and noise signals, as well as with varying speaker and microphone positions. Due to the propagation delay and spectral coloration from the loudspeaker and microphone, exact target signals cannot be accurately defined in a real-world experiment. Thus, we indirectly evaluated the speech enhancement performance from the speech recognition performance (WER) using the Wav2Vec2.0 model pre-trained with 960 hours of Librispeech.
%We used a model trained on a simulated DNS challenge dataset to perform speech enhancement on the signals measured in real-world environments and evaluated the performance by comparing the enhanced speech through listening and spectrogram analysis. Since the room had reverberation and ground truth measurements were not possible, we did not evaluate the performance using metrics such as SI-SDR. We conducted experiments for various speaker positions and microphone positions to assess the applicability of the model in real-world scenarios. Through these experiments in real environments, we validated the model's potential for real-world applications. The SNR is set to -5\,dB and the reverberation time is approximately 0.4-0.6 seconds.

\begin{table}
\caption{Results on real-world experiments (4CH)\label{tab:realexp}}
\renewcommand{\arraystretch}{1.5}
  \centering
  \begin{tabularx}{8cm}{>{\centering\arraybackslash}X||>{\centering\arraybackslash}X|>{\centering\arraybackslash}X|>{\centering\arraybackslash}X}
    \hline
     & Unprocessed & Enhanced & Clean \\
    \hline\hline
    WER (\%)$\downarrow$ & 53.3 & 5.2 & 3.7 \\
    \hline
  \end{tabularx}
\end{table}

\begin{figure}
\centering
\includegraphics[width=3.7in]{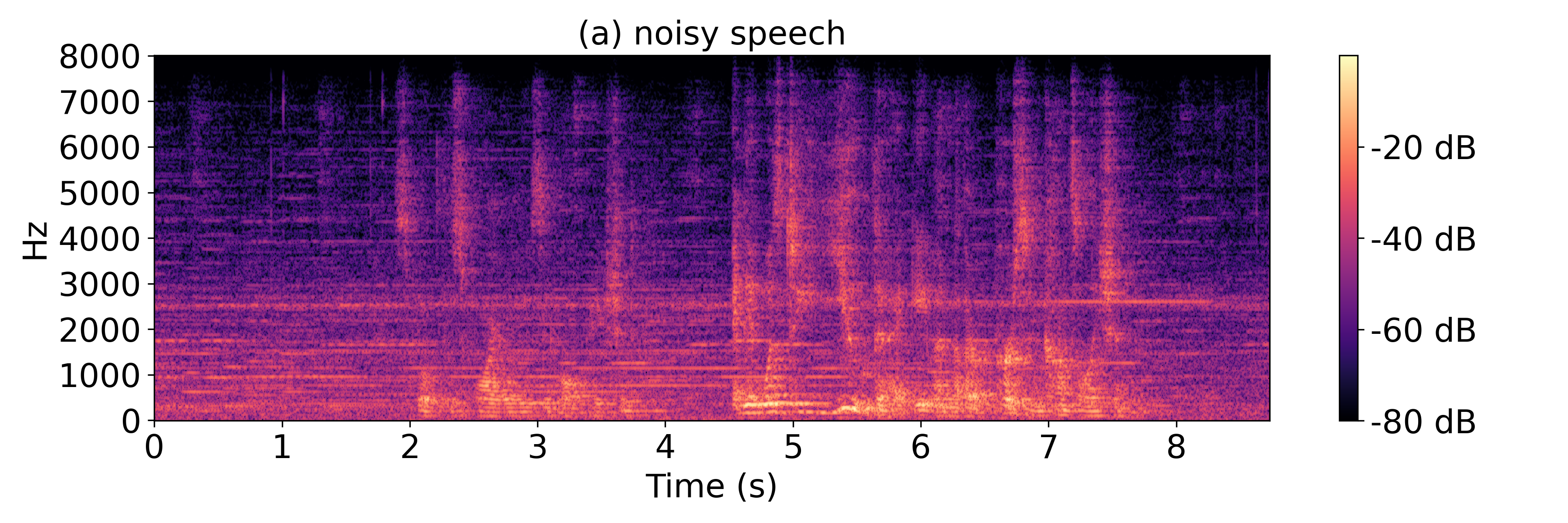}
\includegraphics[width=3.7in]{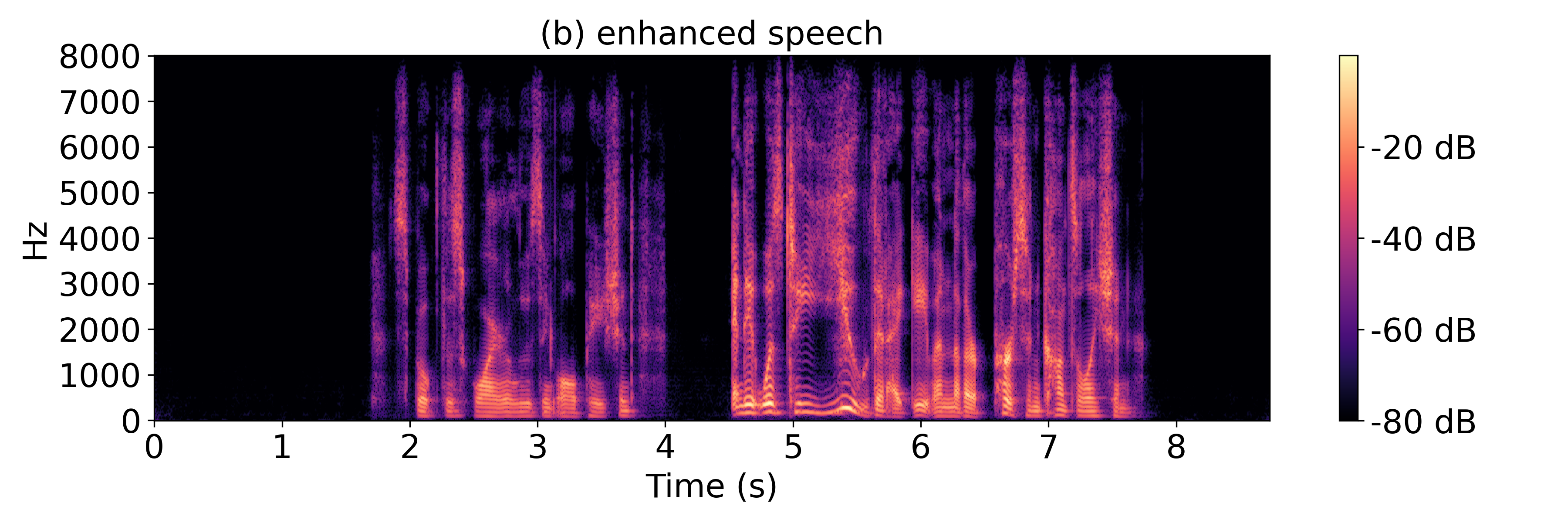}
\includegraphics[width=3.7in]{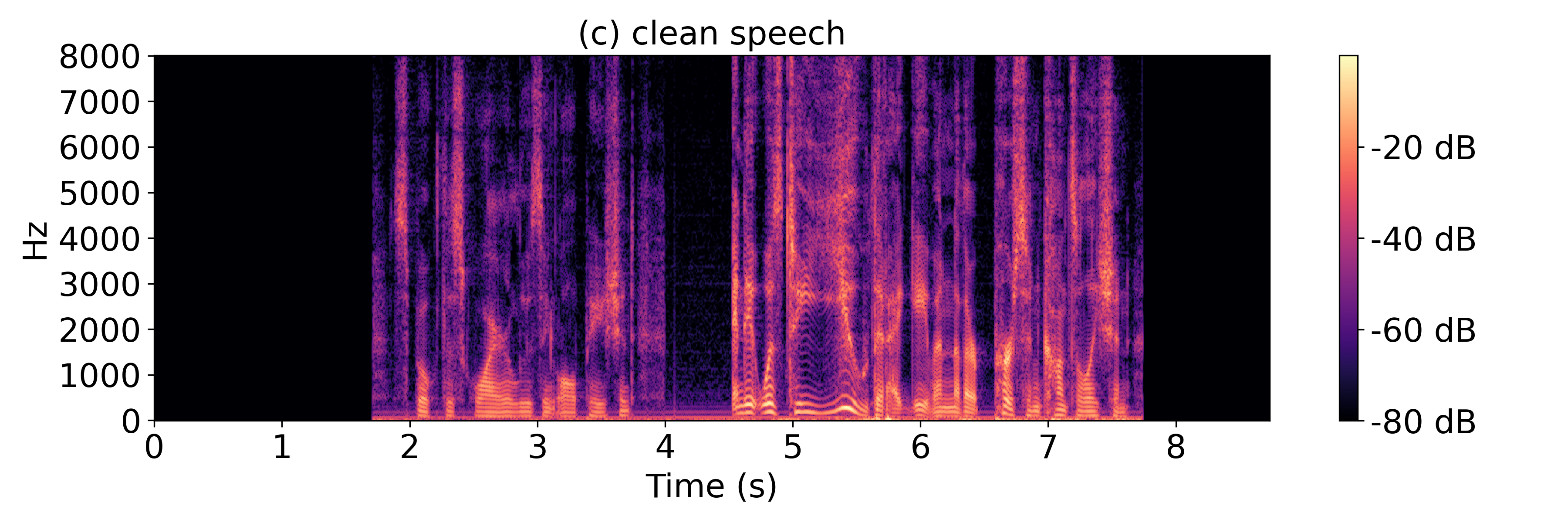}
\caption{Spectrogram example of real-world environment experiment, (a) noisy reverberant speech, (b) enhanced speech, (c) clean speech}
\label{fig:realexp}
\end{figure}

Table\,\ref{tab:realexp} presents the WERs for unprocessed noisy reverberant speech, enhanced speech, and clean speech, which are 53.3\%, 5.2\%, and 3.7\%, respectively. While the WER for enhanced speech is not as low as in the test with the simulated L3DAS22 dataset, it is still remarkably close to that of clean speech, considering the high level of noise and potential bandwidth limitation by loudspeakers and microphones. During unofficial listening tests of the enhanced audio clips, the noise and reverberation were barely perceptible, and the speech sounded as clear as a clean signal. %The speech is intelligible and its content can be sufficiently understood in terms of both perceptual quality and intelligibility. 
Fig.\,\ref{fig:realexp} shows the spectrograms of the noisy reverberant, enhanced, and clean speeches. It is evident that both noise and reverberation are significantly suppressed, leaving only clean speech, and the formants in the high-frequency range are well restored. %The experimental setup was conducted in an empty office, rather than an acoustically designed room. 
The successful speech enhancement in the unseen real-world environment demonstrates that the model can adapt effectively to unseen environments and be utilized in practical scenarios without significant performance degradation. %operates adaptively in unseen environments, examining its usability in real-world scenarios. 
Moreover, the model trained on a circular array with a radius of 10\,cm performed well on the unseen real array with a smaller radius. Although further comprehensive studies are required, this preliminary observation suggests that the proposed model exhibits some flexibility in accommodating different array geometries.

\section{Conclusion}
We proposed DeFTAN-II to address the challenges associated with handling local relationships, high computational complexity, and inefficient memory usage of transformer-based models for multichannel speech enhancement tasks. Our proposed model combines the CNN and transformer architectures to effectively capture relationships between channel, frequency, and time. The subgroup processing incorporated in the CEA and DPFN realized the more efficient transformer architecture by combining features locally emphasized by CNN with conventional ones. 
Furthermore, we designed a network that efficiently extracts spatial features within limited channel dimensions. To optimize the trade-off between the performance and computational cost in relation to the feature map size, we sequentially compressed essential features using the SDB. Iterative utilization of the SDB facilitates a more precise mapping by aggregating the channel and local time-frequency relations. Through a series of comprehensive experiments, we thoroughly evaluated and validated the performance and scalability of our proposed model. 
%By training and evaluating various datasets, 
Comparative assessments against state-of-the-art models clearly demonstrate the superiority of our approach, and consistent performance gain obtained over various datasets shows that the model exhibits negligible data bias. Additionally, by evaluating the pre-trained model on real-environment recordings, we examined the model's applicability in real-world scenarios.

\section*{Acknowledgments}
This work was supported by the BK21 Four program through the National Research Foundation (NRF) funded by the Ministry of Education of Korea, the National Research Council of Science and Technology (NST) granted by the Korean government (MSIT)(No. CRC21011), and the Center for Applied Research in Artificial Intelligence (CARAI) funded by DAPA and ADD (UD230017TD).

 \appendix[Complexity analysis of subgroup processing]
This section presents the analysis of the computational complexity and memory usage of SDB, CEA, and DPFN, which are utilized to derive the computational complexities and memory usages listed in Tables\,\ref{tab:DenseCost}, \ref{tab:AttCost}, and \ref{tab:FFWcost}.

The computational complexities of the 1D and 2D convolution layers with kernel sizes $k$ and $(k, k)$, respectively, are given by \cite{hassani2023neighborhood},
\begin{align}
\label{eq:timecomplex}
\begin{aligned}
T_{conv, 1D} &= C_{in}C_{out}kL,~~ 
T_{conv, 2D} = C_{in}C_{out}k^{2}TF
\end{aligned}
\end{align}
where $C_{in}$ and $C_{out}$ indicate the channel dimensions of the input and output tensors, respectively. 
%In this study, we do not consider cases where the length of the sequence dimension ($T$ or $F$ or $L$) changes due to any convolution layer.

% \begin{figure}
% \centering
% \includegraphics[width=3.5in]{Figure/Attention.png}
% \caption{(a) Scaled dot-product attention, (b) Multi-head self-attention}
% \label{fig:Att}
% \end{figure}

Next, the memory usages, $S_{conv,2D}$ and $S_{conv,1D}$, of 1D and 2D convolution layers are given by\cite{hassani2023neighborhood}
\begin{align}
\begin{aligned}
S_{conv,1D} = C_{in}C_{out}k, ~~ S_{conv,2D} = C_{in}C_{out}k^2\\
\end{aligned}
\end{align}
The biases in convolution layers occupy much lesser memory than that of the convolution kernels, and therefore memory occupancy is neglected in this paper.  

\subsection{Complexity analysis of split dense block}
%, and other symbols follow the notations used in Table\,\ref{tab:DenseCost}. %definitions in Table\,\ref{tab:DenseCost}.
\subsubsection{Dense block}
A dense block concatenates the input and output of the preceding convolutional layer and uses it as the input to the subsequent convolutional layer, repeating the process a total of $G$ times. The number of input and output channels ($C_{in}, C_{out})$ in the $g$-th convolution layer is given by $gC$ and $C$, respectively; therefore, the total computational complexity $T_{dense}$ and memory usage $S_{dense}$ of the dense block correspond to
\begin{align}
\begin{aligned}
T_{dense} &= \sum\limits_{g = 1}^G {gC\cdot C{k^2}TF} = \frac{{G(G + 1)}}{2}{C^2}{k^2}TF\\
S_{dense} &= \sum\limits_{g = 1}^G {gC \cdot Ck^2} = \frac{{G(G + 1)}}{2}{C^2}{k^2}.
\end{aligned}
\end{align}

\subsubsection{Group dense block}
The group dense block employs group convolutions using $G$ groups with channel dimensions reduced by $G$ times. Accordingly, the computational complexity $T_{group}$ and memory usage $S_{group}$ of the group dense block can be reduced to
%is designed to alleviate the complexity of dense blocks by employing group convolutions with $G$ groups.  
\begin{align}
\begin{aligned}
T_{group} &= G\sum\limits_{g = 1}^G {\frac{{gC}}{G}\cdot \frac{C}{G}{k^2}TF} = \frac{{G + 1}}{2}{C^2}{k^2}TF\\
S_{group} &= G\sum\limits_{g = 1}^G {\frac{{gC}}{G}\cdot \frac{C}{G}{k^2}} = \frac{{G + 1}}{2}{C^2}{k^2}.
\end{aligned}
\end{align}

\subsubsection{2D Split Dense Block (SDB)}
In 2D SDB, the channel dimensions of the input and output are equal to $(C/G, C/G)$ for the first layer, and $(2C/G$, $C/G)$ for the remaining layers. Therefore, the computational complexity $T_{2D}$ and memory usage $S_{2D}$ of 2D SDB can be written as
\begin{align}
\begin{aligned}
T_{2D} &= \left(\frac{C}{G}\cdot \frac{C}{G} + \sum\limits_{g = 2}^G{\frac{2C}{G}}\cdot \frac{C}{G}\right){k^2}TF = \frac{{2G - 1}}{{{G^2}}}{C^2}{k^2}TF\\
S_{2D} &= \left(\frac{C}{G}\cdot \frac{C}{G} + \sum\limits_{g = 2}^G{\frac{2C}{G}}\cdot \frac{C}{G}\right){k^2} = \frac{{2G - 1}}{{{G^2}}}{C^2}{k^2}.
\end{aligned}
\end{align}

\subsubsection{1D Split Dense Block (SDB)}
Similar to 2D SDB, the computational complexity $T_{1D}$ and memory usage $S_{1D}$ of 1D SDB can be obtained as
\begin{align}
\begin{aligned}
T_{1D} &= \frac{{2G - 1}}{{{G^2}}}{C^2}{k^2}L,\\
S_{1D} &= \frac{{2G - 1}}{{{G^2}}}{C^2}{k^2},
\end{aligned}
\end{align}
where $L$ denotes the length of the sequence dimension ($T$ or $F$).

\subsection{Complexity analysis of convolutional efficient attention}
In this section, $D$ denotes the number of input channels for the attention layer. %, and other symbols follow the definition in Table\,\ref{tab:AttCost}.
\subsubsection{Vanilla attention}
The computational complexity of the scaled dot-product attention is the sum of the complexity involved with the matrix multiplication corresponding to the scaled dot-product attention and that for the linear projections to obtain the query, key, and value. For a single attention head and input/output sizes of $(D, L)$, the sum can be written as
%using (\ref{eq:timecomplex}).
%Fig.\,\ref{fig:Att} depicts the diagram of scaled dot-product attention and multi-head self-attention with $h$ heads. For simplicity, we compare the complexity of a single head. First, the computational complexity of the vanilla self-attention can be computed for the input and output feature map sizes $(D, L)$ as % follows,
\begin{align}
\label{eq:timeatt}
\begin{aligned}
T_{att} &= 2DL^2+3D^2L.
\end{aligned}
\end{align}
By contrast, the memory usage of the vanilla attention layer can be computed as \cite{hassani2023neighborhood}
\begin{align}
\label{eq:spaceatt}
\begin{aligned}
S_{att} &= L^2 + 3D^2,
\end{aligned}
\end{align}
where $L^2$ corresponds to the size of the attention map $L \times L$, and $3D^2$ corresponds to the parameter size of linear projections to obtain the query, key, and value.

%Here, the first and second terms correspond to the computational complexities of the matrix multiplication involved in the scaled dot-product attention and the linear projections to obtain query, key, and value, respectively.
% From (\ref{eq:timeatt}) and (\ref{eq:spaceatt}), the computational complexity $T_{att}$ and memory usage $S_{att}$ of the vanilla attention can be written as
% \begin{align}
% \begin{aligned}
% T_{att} &= 2DL^2 + 3D^2L,\\
% S_{att} &= L^2 + 3D^2
% \end{aligned}
% \end{align}

\subsubsection{Efficient Attention (EA)}
In EA, the computational complexity of scaled dot-product attention is reduced by multiplying $\mathbf{K}^T$ and $\mathbf{V}$, instead of $\mathbf{Q}$ and $\mathbf{K}^T$. Accordingly, the computational complexity of EA can be written as
\begin{align}
\begin{aligned}
T_{EA} &= 2D^2L + 3D^2L = 5D^2L
\end{aligned}
\end{align}
 %the computational complexity of scaled dot-product attention, $2DL^2$, is reduced to $2D^2L$.
The memory usage of EA $S_{EA}$ is also reduced owing to the decreased size of the attention map ($D^2$), that is,
\begin{align}
\begin{aligned}
S_{EA} &= D^2 + 3D^2 = 4D^2.
\end{aligned}
\end{align}
%Similar to the computational complexity of EA, the memory usage of EA is also reduced due to the smaller size of the attention map, which is $D \times D$ instead of $L \times L$.

\subsubsection{Convolutional Efficient Attention (CEA)}
The computational complexity $T_{CEA}$ and memory usage $S_{CEA}$ of CEA can be obtained as
\begin{align}
\begin{aligned}
T_{CEA} &= D\cdot 2Dk^{2}L + T_{EA} = (5+2k^2)D^2L \\
S_{CEA} &= D\cdot 2Dk^{2} + S_{EA} = (4+2k^2)D^2,
\end{aligned}
\end{align}
where the first terms of $T_{CEA}$ and $S_{CEA}$ correspond to the computational complexity and memory usage of the additional 1D Conv with a kernel size of $k$. %CEA exhibits higher computational complexity and memory usage compared to EA due to the additional convolution. However, it still remains lower in both metrics than vanilla attention.

\subsection{Complexity analysis of dual-path feedforward network}
%In this section, $D$ denotes the number of the channels of the input and output of FFW, and other symbols follow the definition in Table\,\ref{tab:FFWcost}.
\subsubsection{Vanilla Feedforward Network (FFW)}
The vanilla feedforward network (FFW) can be described by the following equation:
\begin{align}
\begin{aligned}
FFW(\mathbf{X}) &= \mathrm{W}_{o}(\mathrm{GELU}(\mathrm{W}_{i}(\mathbf{X}))),
\end{aligned}
\end{align}
%where $\mathrm{W}_{i}$, $\mathrm{W}_{o}$ denote the pointwise convolutions applied to the features.
The number of channels increased owing to linear projection $\mathrm{W}_{i}$ is $4D$.
Therefore, the computational complexity $T_{FFW}$ and memory usage $S_{FFW}$ of the vanilla FFW can be written as
\begin{align}
\begin{aligned}
T_{FFW} &= D\cdot 4DL + 4D\cdot DL = 8D^2L \\
S_{FFW} &= D\cdot 4D + 4D\cdot D = 8D^2,
\end{aligned}
\end{align}
where the first term and second term of $T_{FFW}$, $S_{FFW}$ correspond to the computational complexity and memory usage of $\mathrm{W}_{i}$ and $\mathrm{W}_{o}$, respectively.

\subsubsection{Depthwise Convolutional Feedforward Networks (DCFN)}
DCFN utilizes depthwise convolution of kernel size $l$ for all features in FFW. Therefore, the overhead corresponding to depthwise convolution is added to its computational complexity and memory usage as
\begin{align}
\begin{aligned}
T_{DCFN} &= T_{FFW} + 4Dl^2L = 8(1 + \frac{l^2}{2D})D^2L \\
S_{DCFN} &= S_{FFW} + 4Dl^2 = 8(1+\frac{l^2}{2D})D^2.
\end{aligned}
\end{align}

\subsubsection{Dual-Path Feedforward Networks (DPFN)}
DPFN $T_{DPFN}$ applies dilated convolution only to half of the features ($2D$ channels); therefore, its complexity and memory usage can be written as
\begin{align}
\begin{aligned}
T_{DPFN} &= T_{FFW} + 2D\cdot 2Dl^2L = 8(1+\frac{l^2}{2})D^2L \\
S_{DPFN} &= S_{FFW} + 2D\cdot 2Dl^2 = 8(1+\frac{l^2}{2})D^2.
\end{aligned}
\end{align}
%where the second term of $T_{DPFN}$ and $S_{DPFN}$ correspond to the computational complexity and memory usage of convolution of subgroup processing.

\subsubsection{Convolutional Feedforward Networks (CFN)}
When CFN applies convolution to all features, the complexity and memory usage are given by
\begin{align}
\begin{aligned}
T_{CFN} &= T_{FFW} + 4D\cdot 4Dl^2L = 8(1+2l^2)D^2L \\
S_{CFN} &= S_{FFW} + 4D\cdot 4Dl^2 = 8(1+2l^2)D^2.
\end{aligned}
\end{align}

\bibliography{ref}

\begin{thebibliography}{10}
\providecommand{\url}[1]{#1}
\csname url@samestyle\endcsname
\providecommand{\newblock}{\relax}
\providecommand{\bibinfo}[2]{#2}
\providecommand{\BIBentrySTDinterwordspacing}{\spaceskip=0pt\relax}
\providecommand{\BIBentryALTinterwordstretchfactor}{4}
\providecommand{\BIBentryALTinterwordspacing}{\spaceskip=\fontdimen2\font plus
\BIBentryALTinterwordstretchfactor\fontdimen3\font minus
  \fontdimen4\font\relax}
\providecommand{\BIBforeignlanguage}[2]{{%
\expandafter\ifx\csname l@#1\endcsname\relax
\typeout{** WARNING: IEEEtran.bst: No hyphenation pattern has been}%
\typeout{** loaded for the language `#1'. Using the pattern for}%
\typeout{** the default language instead.}%
\else
\language=\csname l@#1\endcsname
\fi
#2}}
\providecommand{\BIBdecl}{\relax}
\BIBdecl

\bibitem{donley2021easycom}
J.~Donley \emph{et~al.}, ``Easy{C}om: An augmented reality dataset to support
  algorithms for easy communication in noisy environments,'' \emph{arXiv
  preprint arXiv:2107.04174}, 2021.

\bibitem{guiraud2022introduction}
P.~Guiraud, S.~Hafezi, P.~A. Naylor, A.~H. Moore, J.~Donley, V.~Tourbabin, and
  T.~Lunner, ``An introduction to the speech enhancement for augmented reality
  (spear) challenge,'' in \emph{Proc. Int. Workshop. Acoust. Sig. Enhancement},
  Bamberg, Germany, 2022, pp. 1--5.

\bibitem{yoshioka2015ntt}
T.~Yoshioka \emph{et~al.}, ``The {NTT} {CHiME}-3 system: Advances in speech
  enhancement and recognition for mobile multi-microphone devices,'' in
  \emph{2015 IEEE Workshop on Automat. Speech Recognit. Understanding},
  Scottsdale, AZ, 2015, pp. 436--443.

\bibitem{ren2021causal}
X.~Ren \emph{et~al.}, ``A causal u-net based neural beamforming network for
  real-time multi-channel speech enhancement.'' in \emph{Proc. Interspeech},
  Brno, Czechia, 2021, pp. 1832--1836.

\bibitem{pandey2022tparn}
A.~Pandey, B.~Xu, A.~Kumar, J.~Donley, P.~Calamia, and D.~Wang, ``{TPARN}:
  Triple-path attentive recurrent network for time-domain multichannel speech
  enhancement,'' in \emph{Proc. IEEE Int. Conf. Acoust., Speech, Signal
  Process.}, Singapore, 2022, pp. 6497--6501.

\bibitem{tolooshams2020channel}
B.~Tolooshams, R.~Giri, A.~H. Song, U.~Isik, and A.~Krishnaswamy,
  ``Channel-attention dense u-net for multichannel speech enhancement,'' in
  \emph{Proc. IEEE Int. Conf. Acoust., Speech, Signal Process.}, Barcelona,
  Spain, 2020, pp. 836--840.

\bibitem{wang2020complex}
Z.-Q. Wang, P.~Wang, and D.~Wang, ``Complex spectral mapping for single-and
  multi-channel speech enhancement and robust {ASR},'' \emph{IEEE/ACM Trans.
  Audio, Speech, Lang. Process.}, vol.~28, pp. 1778--1787, May. 2020.

\bibitem{wang2020multi}
Z.-Q. Wang and D.~Wang, ``Multi-microphone complex spectral mapping for speech
  dereverberation,'' in \emph{Proc. IEEE Int. Conf. Acoust., Speech, Signal
  Process.}, Barcelona, Spain, 2020, pp. 486--490.

\bibitem{tesch2022insights}
K.~Tesch and T.~Gerkmann, ``Insights into deep non-linear filters for improved
  multi-channel speech enhancement,'' \emph{IEEE/ACM Trans. Audio, Speech,
  Lang. Process.}, vol.~31, pp. 563--575, Nov. 2022.

\bibitem{li2022embedding}
A.~Li, W.~Liu, C.~Zheng, and X.~Li, ``Embedding and beamforming: All-neural
  causal beamformer for multichannel speech enhancement,'' in \emph{Proc. IEEE
  Int. Conf. Acoust., Speech, Signal Process.}, Singapore, 2022, pp.
  6487--6491.

\bibitem{pandey2022multichannel}
A.~Pandey, B.~Xu, A.~Kumar, J.~Donley, P.~Calamia, and D.~Wang, ``Multichannel
  speech enhancement without beamforming,'' in \emph{Proc. IEEE Int. Conf.
  Acoust., Speech, Signal Process.}, Singapore, 2022, pp. 6502--6506.

\bibitem{liu2022drc}
J.~Liu and X.~Zhang, ``{DRC-NET}: Densely connected recurrent convolutional
  neural network for speech dereverberation,'' in \emph{Proc. IEEE Int. Conf.
  Acoust., Speech, Signal Process.}, Singapore, 2022, pp. 166--170.

\bibitem{shubo2023spatial}
L.~Shubo \emph{et~al.}, ``{Spatial-DCCRN}: {DCCRN} equipped with frame-level
  angle feature and hybrid filtering for multi-channel speech enhancement,'' in
  \emph{Proc. IEEE Spoken Lang. Technol. Workshop}, Doha, Qatar, 2023, pp.
  436--443.

\bibitem{yang2023mcnet}
Y.~Yang, C.~Quan, and X.~Li, ``{McNet}: Fuse multiple cues for multichannel
  speech enhancement,'' in \emph{Proc. IEEE Int. Conf. Acoust., Speech, Signal
  Process.}, Rhodes, Greece, 2023, pp. 1--5.

\bibitem{lee2023deft}
D.~Lee and J.-W. Choi, ``De{FT-AN}: Dense frequency-time attentive network for
  multichannel speech enhancement,'' \emph{IEEE Signal Proc. Lett.}, vol.~30,
  pp. 155--159, Feb. 2023.

\bibitem{xu2013experimental}
Y.~Xu, J.~Du, L.-R. Dai, and C.-H. Lee, ``An experimental study on speech
  enhancement based on deep neural networks,'' \emph{IEEE Signal Proc. Lett.},
  vol.~21, no.~1, pp. 65--68, Nov. 2013.

\bibitem{lu2013speech}
X.~Lu, Y.~Tsao, S.~Matsuda, and C.~Hori, ``Speech enhancement based on deep
  denoising autoencoder.'' in \emph{Proc. Interspeech}, vol. 2013, Lyon,
  France, 2013, pp. 436--440.

\bibitem{xu2014regression}
Y.~Xu, J.~Du, L.-R. Dai, and C.-H. Lee, ``A regression approach to speech
  enhancement based on deep neural networks,'' \emph{IEEE/ACM Trans. Audio,
  Speech, Lang. Process.}, vol.~23, no.~1, pp. 7--19, Oct. 2014.

\bibitem{isik2016single}
Y.~Isik, J.~Le~Roux, Z.~Chen, S.~Watanabe, and J.~R. Hershey, ``Single-channel
  multi-speaker separation using deep clustering,'' in \emph{Proc.
  Interspeech}, San Francisco, CA, 2016, pp. 545--549.

\bibitem{williamson2015complex}
D.~S. Williamson, Y.~Wang, and D.~Wang, ``Complex ratio masking for monaural
  speech separation,'' \emph{IEEE/ACM Trans. Audio, Speech, Lang. Process.},
  vol.~24, no.~3, pp. 483--492, Dec. 2015.

\bibitem{park2017fully}
S.~R. Park and J.~W. Lee, ``A fully convolutional neural network for speech
  enhancement,'' in \emph{Proc. Interspeech 2017}, Stockholm, Sweden, 2017, pp.
  1993--1997.

\bibitem{wang2018supervised}
D.~Wang and J.~Chen, ``Supervised speech separation based on deep learning: An
  overview,'' \emph{IEEE/ACM Trans. Audio, Speech, Lang. Process.}, vol.~26,
  no.~10, pp. 1702--1726, May. 2018.

\bibitem{fu2017raw}
S.-W. Fu, Y.~Tsao, X.~Lu, and H.~Kawai, ``Raw waveform-based speech enhancement
  by fully convolutional networks,'' in \emph{Proc. Asia-Pacific Signal, Info.
  Process. Assoc. Annu. Summit Conf.}, 2017, pp. 006--012.

\bibitem{luo2018tasnet}
Y.~Luo and N.~Mesgarani, ``{TasNet}: time-domain audio separation network for
  real-time, single-channel speech separation,'' in \emph{Proc. IEEE Int. Conf.
  Acoust., Speech, Signal Process.}, Calgary, AB, Canada, 2018, pp. 696--700.

\bibitem{venkataramani2018end}
S.~Venkataramani, J.~Casebeer, and P.~Smaragdis, ``End-to-end source separation
  with adaptive front-ends,'' in \emph{Proc. 52nd Asilomar Conf. Signals, Sys.,
  and Comput.}, 2018, pp. 684--688.

\bibitem{luo2019conv}
Y.~Luo and N.~Mesgarani, ``{Conv-TasNet}: Surpassing ideal time--frequency
  magnitude masking for speech separation,'' \emph{IEEE/ACM Trans. Audio,
  Speech, Lang. Process.}, vol.~27, no.~8, pp. 1256--1266, May. 2019.

\bibitem{rixen2022sfsrnet}
J.~Rixen and M.~Renz, ``{SFSRNet}: Super-resolution for single-channel audio
  source separation,'' in \emph{Proceedings of the AAAI}, vol.~36, no.~10,
  Virtual, 2022, pp. 11\,220--11\,228.

\bibitem{tzinis2020sudo}
E.~Tzinis, Z.~Wang, and P.~Smaragdis, ``Sudo rm-rf: Efficient networks for
  universal audio source separation,'' in \emph{Proc. IEEE 30th Int. Workshop
  Mach. Learn. Signal Process.}, Espoo, Finland, 2020, pp. 1--6.

\bibitem{rixen2022qdpn}
J.~Rixen and M.~Renz, ``{QDPN}-quasi-dual-path network for single-channel
  speech separation,'' in \emph{Proc. Interspeech}, Incheon, Korea, 2022, pp.
  5353--5357.

\bibitem{luo2020dual}
Y.~Luo, Z.~Chen, and T.~Yoshioka, ``Dual-path {RNN}: efficient long sequence
  modeling for time-domain single-channel speech separation,'' in \emph{Proc.
  IEEE Int. Conf. Acoust., Speech, Signal Process.}, 2020, pp. 46--50.

\bibitem{fu2017complex}
S.-W. Fu, T.-y. Hu, Y.~Tsao, and X.~Lu, ``Complex spectrogram enhancement by
  convolutional neural network with multi-metrics learning,'' in \emph{Proc.
  IEEE 30th Int. Workshop Mach. Learn. Signal Process.}, Tokyo, Japan, 2017,
  pp. 1--6.

\bibitem{tan2019complex}
K.~Tan and D.~Wang, ``Complex spectral mapping with a convolutional recurrent
  network for monaural speech enhancement,'' in \emph{Proc. IEEE Int. Conf.
  Acoust., Speech, Signal Process.}, Brighton, UK, 2019, pp. 6865--6869.

\bibitem{pandey2020densely}
A.~Pandey and D.~Wang, ``Densely connected neural network with dilated
  convolutions for real-time speech enhancement in the time domain,'' in
  \emph{Proc. IEEE Int. Conf. Acoust., Speech, Signal Process.}, Apr. 2020, pp.
  6629--6633.

\bibitem{yu2022dual}
G.~Yu, A.~Li, C.~Zheng, Y.~Guo, Y.~Wang, and H.~Wang, ``Dual-branch
  attention-in-attention transformer for single-channel speech enhancement,''
  in \emph{Proc. IEEE Int. Conf. Acoust., Speech, Signal Process.}, Singapore,
  2022, pp. 7847--7851.

\bibitem{pandey2021dense}
A.~Pandey and D.~Wang, ``Dense {CNN} with self-attention for time-domain speech
  enhancement,'' \emph{IEEE/ACM Trans. Audio, Speech, Lang. Process.}, vol.~29,
  pp. 1270--1279, Mar. 2021.

\bibitem{yang2022tfpsnet}
L.~Yang, W.~Liu, and W.~Wang, ``{TFPSNet}: Time-frequency domain path scanning
  network for speech separation,'' in \emph{Proc. IEEE Int. Conf. Acoust.,
  Speech, Signal Process.}, 2022, pp. 6842--6846.

\bibitem{wang2023tf}
Z.-Q. Wang, S.~Cornell, S.~Choi, Y.~Lee, B.-Y. Kim, and S.~Watanabe,
  ``{TF-GridNet}: Making time-frequency domain models great again for monaural
  speaker separation,'' in \emph{Proc. IEEE Int. Conf. Acoust., Speech, Signal
  Process.}, Rhodes, Greece, 2023, pp. 1--5.

\bibitem{elman1990finding}
J.~L. Elman, ``Finding structure in time,'' \emph{Cognitive science}, vol.~14,
  no.~2, pp. 179--211, Mar. 1990.

\bibitem{lecun1998gradient}
Y.~LeCun, L.~Bottou, Y.~Bengio, and P.~Haffner, ``Gradient-based learning
  applied to document recognition,'' \emph{Proc. IEEE}, vol.~86, no.~11, pp.
  2278--2324, Nov. 1998.

\bibitem{vaswani2017attention}
A.~Vaswani, N.~Shazeer, N.~Parmar, J.~Uszkoreit, L.~Jones, A.~N. Gomez,
  {\L}.~Kaiser, and I.~Polosukhin, ``Attention is all you need,'' \emph{Proc.
  Advances in Neural Info. Process. Sys.}, vol.~30, Long Beach, CA, 2017.

\bibitem{chen2015speech}
Z.~Chen, S.~Watanabe, H.~Erdogan, and J.~R. Hershey, ``Speech enhancement and
  recognition using multi-task learning of long short-term memory recurrent
  neural networks,'' in \emph{16th Annu. Conf. Int. Speech. Commun. Assoc.},
  2015.

\bibitem{weninger2015speech}
F.~Weninger \emph{et~al.}, ``Speech enhancement with {LSTM} recurrent neural
  networks and its application to noise-robust {ASR},'' in \emph{Int. Conf.
  Latent Variable Anal., Signal Separation}.\hskip 1em plus 0.5em minus
  0.4em\relax Springer, Liberec, Czech, 2015, pp. 91--99.

\bibitem{hochreiter1997long}
S.~Hochreiter and J.~Schmidhuber, ``Long short-term memory,'' \emph{Neural
  computation}, vol.~9, no.~8, pp. 1735--1780, 1997.

\bibitem{subakan2021attention}
C.~Subakan, M.~Ravanelli, S.~Cornell, M.~Bronzi, and J.~Zhong, ``Attention is
  all you need in speech separation,'' in \emph{Proc. IEEE Int. Conf. Acoust.,
  Speech, Signal Process.}, Virtual, 2021, pp. 21--25.

\bibitem{chen2020dual}
J.~Chen, Q.~Mao, and D.~Liu, ``Dual-path transformer network: Direct
  context-aware modeling for end-to-end monaural speech separation,''
  \emph{Proc. Interspeech}, pp. 2642--2646, Shanghai, China, 2020.

\bibitem{huang2017densely}
G.~Huang, Z.~Liu, L.~Van Der~Maaten, and K.~Q. Weinberger, ``Densely connected
  convolutional networks,'' in \emph{Proc. IEEE Conf. Comp. Vis. Pattern
  Recognit.}, Honolulu, HI, 2017, pp. 4700--4708.

\bibitem{wang2022tf}
Z.-Q. Wang, S.~Cornell, S.~Choi, Y.~Lee, B.-Y. Kim, and S.~Watanabe,
  ``{TF-GridNet}: Integrating full-and sub-band modeling for speech
  separation,'' \emph{arXiv preprint arXiv:2211.12433}, 2022.

\bibitem{gulati2020conformer}
A.~Gulati \emph{et~al.}, ``Conformer: Convolution-augmented transformer for
  speech recognition,'' \emph{Proc. Interspeech}, pp. 5036--5040, Shanghai,
  China, 2020.

\bibitem{guo2022cmt}
J.~Guo, K.~Han, H.~Wu, Y.~Tang, X.~Chen, Y.~Wang, and C.~Xu, ``{CMT}:
  Convolutional neural networks meet vision transformers,'' in \emph{Proc.
  IEEE/CVF Conf. Comp. Vis. Pattern Recognit.}, New Orleans, LA, 2022, pp.
  12\,175--12\,185.

\bibitem{liu2021swin}
Z.~Liu, Y.~Lin, Y.~Cao, H.~Hu, Y.~Wei, Z.~Zhang, S.~Lin, and B.~Guo, ``Swin
  transformer: Hierarchical vision transformer using shifted windows,'' in
  \emph{Proc. IEEE/CVF Int. Conf. Comp. Vis.}, Virtual, 2021, pp.
  10\,012--10\,022.

\bibitem{li2019enhancing}
S.~Li, X.~Jin, Y.~Xuan, X.~Zhou, W.~Chen, Y.-X. Wang, and X.~Yan, ``Enhancing
  the locality and breaking the memory bottleneck of transformer on time series
  forecasting,'' \emph{Advances in Neural Info. Process. Sys.}, vol.~32,
  Vancouver, BC, Canada, 2019.

\bibitem{guizzo2022l3das22}
E.~Guizzo, C.~Marinoni, M.~Pennese, X.~Ren, X.~Zheng, C.~Zhang, B.~Masiero,
  A.~Uncini, and D.~Comminiello, ``{L3DAS22} challenge: Learning 3d audio
  sources in a real office environment,'' in \emph{Proc. IEEE Int. Conf.
  Acoust., Speech, Signal Process.}\hskip 1em plus 0.5em minus 0.4em\relax
  IEEE, Singapore, 2022, pp. 9186--9190.

\bibitem{he2015delving}
K.~He, X.~Zhang, S.~Ren, and J.~Sun, ``Delving deep into rectifiers: Surpassing
  human-level performance on imagenet classification,'' in \emph{Proc. IEEE
  Int. Conf. Comp. Vis.}, Santiago, Chile, 2015, pp. 1026--1034.

\bibitem{hassani2023neighborhood}
A.~Hassani, S.~Walton, J.~Li, S.~Li, and H.~Shi, ``Neighborhood attention
  transformer,'' in \emph{Proc. IEEE/CVF Conf. Comp. Vis. Pattern Recognit.},
  Vancouver, BC, Canada, 2023, pp. 6185--6194.

\bibitem{shen2021efficient}
Z.~Shen, M.~Zhang, H.~Zhao, S.~Yi, and H.~Li, ``Efficient attention: Attention
  with linear complexities,'' in \emph{Proc. IEEE/CVF Winter Conf. Appl. Comp.
  Vis.}, Waikoloa, HI, 2021, pp. 3531--3539.

\bibitem{dauphin2017language}
Y.~N. Dauphin, A.~Fan, M.~Auli, and D.~Grangier, ``Language modeling with gated
  convolutional networks,'' in \emph{Proc. Int. Conf. Mach. Learn.}, Sydney,
  Australia, 2017, pp. 933--941.

\bibitem{koizumi2021df}
Y.~Koizumi \emph{et~al.}, ``{DF-Conformer}: Integrated architecture of
  conv-tasnet and conformer using linear complexity self-attention for speech
  enhancement,'' in \emph{Proc. IEEE Workshop Appl. Signal Process. Audio,
  Acoust.}, New Paltz, NY, 2021, pp. 161--165.

\bibitem{tan2021efficientnetv2}
M.~Tan and Q.~Le, ``{EfficientNetV2}: Smaller models and faster training,'' in
  \emph{Proc. Int. Conf. Mach. Learn.}, Virtual, 2021, pp. 10\,096--10\,106.

\bibitem{robinson1995wsjcamo}
T.~Robinson, J.~Fransen, D.~Pye, J.~Foote, and S.~Renals, ``{WSJCAM0}: a
  british english speech corpus for large vocabulary continuous speech
  recognition,'' in \emph{Proc. IEEE Int. Conf. Acoust., Speech, Signal
  Process.}, vol.~1, Detroit, MI, 1995, pp. 81--84.

\bibitem{scheibler2018pyroomacoustics}
R.~Scheibler, E.~Bezzam, and I.~Dokmani{\'c}, ``Pyroomacoustics: A python
  package for audio room simulation and array processing algorithms,'' in
  \emph{Proc. IEEE Int. Conf. Acoust., Speech, Signal Process.}, Calgary, AB,
  Canada, 2018, pp. 351--355.

\bibitem{kinoshita2016summary}
K.~Kinoshita \emph{et~al.}, ``A summary of the {REVERB} challenge:
  state-of-the-art and remaining challenges in reverberant speech processing
  research,'' \emph{EURASIP J. Adv. Signal Process.}, vol. 2016, pp. 1--19,
  Jan. 2016.

\bibitem{reddy2020interspeech}
C.~K. Reddy \emph{et~al.}, ``The interspeech 2020 deep noise suppression
  challenge: Datasets, subjective testing framework, and challenge results,''
  \emph{Proc. Interspeech}, pp. 2492--2496, 2020.

\bibitem{panayotov2015librispeech}
V.~Panayotov, G.~Chen, D.~Povey, and S.~Khudanpur, ``Librispeech: an {ASR}
  corpus based on public domain audio books,'' in \emph{Proc. IEEE Int. Conf.
  Acoust., Speech, Signal Process.}, South Brisbane, Australia, 2015, pp.
  5206--5210.

\bibitem{fonseca2021fsd50k}
E.~Fonseca, X.~Favory, J.~Pons, F.~Font, and X.~Serra, ``{FSD50k}: an open
  dataset of human-labeled sound events,'' \emph{IEEE/ACM Trans. Audio, Speech,
  Lang. Process.}, vol.~30, pp. 829--852, Dec. 2021.

\bibitem{kingma2014adam}
D.~Kingma, ``Adam: a method for stochastic optimization,'' in \emph{Proc. Int.
  Conf. Learn. Represent.}, Banff, AB, Canada, 2014.

\bibitem{le2019sdr}
J.~Le~Roux, S.~Wisdom, H.~Erdogan, and J.~R. Hershey, ``{SDR}--half-baked or
  well done?'' in \emph{Proc. IEEE Int. Conf. Acoust., Speech, Signal
  Process.}, Brighton, UK, 2019, pp. 626--630.

\bibitem{rix2001perceptual}
A.~W. Rix, J.~G. Beerends, M.~P. Hollier, and A.~P. Hekstra, ``Perceptual
  evaluation of speech quality (pesq)-a new method for speech quality
  assessment of telephone networks and codecs,'' in \emph{Proc. IEEE Int. Conf.
  Acoust., Speech, Signal Process.}, vol.~2, Salt Lake City, UT, 2001, pp.
  749--752.

\bibitem{taal2010short}
C.~H. Taal, R.~C. Hendriks, R.~Heusdens, and J.~Jensen, ``A short-time
  objective intelligibility measure for time-frequency weighted noisy speech,''
  in \emph{Proc. IEEE Int. Conf. Acoust., Speech, Signal Process.}, Dallas, TX,
  2010, pp. 4214--4217.

\bibitem{lu2022towards}
Y.-J. Lu \emph{et~al.}, ``Towards low-distortion multi-channel speech
  enhancement: The {ESPNET-SE} submission to the {L3DAS22} challenge,'' in
  \emph{Proc. IEEE Int. Conf. Acoust., Speech, Signal Process.}, Singapore,
  2022, pp. 9201--9205.

\bibitem{zhang2022multi}
G.~Zhang, L.~Yu, C.~Wang, and J.~Wei, ``Multi-scale temporal frequency
  convolutional network with axial attention for speech enhancement,'' in
  \emph{Proc. IEEE Int. Conf. Acoust., Speech, Signal Process.}, Singapore,
  2022, pp. 9122--9126.

\bibitem{li2022pcg}
J.~Li, Y.~Zhu, D.~Luo, Y.~Liu, G.~Cui, and Z.~Li, ``The {PCG-AIID} system for
  {L3DAS22} challenge: {MIMO} and {MISO} convolutional recurrent network for
  multi channel speech enhancement and speech recognition,'' in \emph{Proc.
  IEEE Int. Conf. Acoust., Speech, Signal Process.}, Singapore, 2022, pp.
  9211--9215.

\bibitem{baevski2020wav2vec}
A.~Baevski, Y.~Zhou, A.~Mohamed, and M.~Auli, ``Wav2vec 2.0: A framework for
  self-supervised learning of speech representations,'' \emph{Advances in
  Neural Info. Process. Sys.}, vol.~33, pp. 12\,449--12\,460, Virtual, 2020.

\end{thebibliography}
\bibliographystyle{IEEEtran}

\end{document}